\documentstyle[12pt,graphicx,subfigure]{article}
\def\rmGeV{~\rm GeV}

\def\Journal#1#2#3#4{{#1} {\bf #2}, #3 (#4)}

\def\NPB{{\em Nucl. Phys.} B}
\def\PLB{{\em Phys. Lett.}  B}
\def\PRL{\em Phys. Rev. Lett.}
\def\PRD{{\em Phys. Rev.} D}
\def\ZPC{{\em Z. Phys.} C}

\def\beq{\begin{equation}}
\def\eeq{\end{equation}}
\def\lsim{\ ^<\llap{$_\sim$}\ }
\def\gsim{\ ^>\llap{$_\sim$}\ }
\def\r2{\sqrt 2}
\def\beq{\begin{equation}}
\def\eeq{\end{equation}}
\def\beqn{\begin{eqnarray}}
\def\eeqn{\end{eqnarray}}

\def\sinW2{\sin^2\theta_W}

\def\mz2{M_{z}^2}
\def\c2b{\cos 2\beta}

\def\mz{M_z}

\def\Fq2{F_{2}(q^2)}

\def\sec2w{sec^2\theta_W}
\def\Omegachi{\Omega_\chi}

\def\tanbeta{{\rm tan}\beta}
\def\gmin2{(g-2)_\mu}

\catcode`\@=11 

\def\lsim{\mathrel{\mathpalette\@versim<}}
\def\gsim{\mathrel{\mathpalette\@versim>}}
\def\@versim#1#2{\vcenter{\offinterlineskip
    \ialign{$\m@th#1\hfil##\hfil$\crcr#2\crcr\sim\crcr } }}

\def\PRL{Phys. Rev. Lett.}

\begin{document}
\begin{flushright}
{TIFR/TH/01-49}\\ 
{NUB-TH/3225}\\
\end{flushright}
\begin{center}
{\Large\bf Supersymmetric Dark Matter 
and Yukawa Unification\\}
\vglue 0.5cm
{Utpal Chattopadhyay$^{(a)}$, Achille Corsetti$^{(b)}$  and 
Pran Nath$^{(b)}$
\vglue 0.2cm
{\em 
$^{(a)}$Department of Theoretical Physics,\\ Tata Institute
of Fundamental Research,Homi Bhabha Road\\
Mumbai 400005, India}\\
{\em $^{(b)}$Department of Physics, Northeastern University, Boston,
MA 02115, USA\\} }
\end{center}
\begin{abstract}

\end{abstract}
An analysis of supersymmetric dark matter under the Yukawa  
unification constraint is given. The analysis utilizes the recently
discovered region of the parameter space of models with gaugino
mass nonuniversalities where large negative supersymmetric corrections to the
b quark mass  appear to allow  $b-\tau$ unification for a positive 
$\mu$ sign consistent 
with the  $b\rightarrow s+\gamma$ and $g_{\mu}-2$ constraints.
In the present analysis we use the revised theoretical determination of 
 $a_{\mu}^{SM}$ ($a_{\mu}=(g_{\mu}-2)/2$) in computing the difference
 $a_{\mu}^{exp}-a_{\mu}^{SM}$  which takes account of a reevaluation of
  the light by light contribution which has a positive sign.
The analysis shows that the region of the
parameter space with nonuniversalities of the gaugino masses which 
 allows for unification of Yukawa couplings also contains
 regions which allow satisfaction of the relic density constraint. 
Specifically we find that the lightest neutralino mass consistent with the relic
 density constraint, $b\tau$ unification for SU(5) and 
 $b-t-\tau$ unification for SO(10) in addition to other constraints lies 
 in the region below $80$ GeV. An analysis of the maximum and the
 minimum neutralino-proton scalar cross section for the allowed
 parameter space including the effect of a new determination
 of the pion-nucleon sigma term is also given. It is  found that 
   the  full parameter space for this class
of models can be explored  in the next generation 
of proposed dark matter detectors.

\section{Introduction}
Recently supersymmetric dark matter has come under a great deal 
of scrutiny due to the fact that the neutralino-proton cross sections 
for a wide class of 
supersymmetric models fall within the range that is accessible
 to the current and planned dark matter 
 experiments\cite{dama,cdms,hdms,genius,cline}. 
Thus some recent studies have
included a variety of effects in the predictions of relic densities
and of detection rates in the direct and in the indirect detection
of dark matter\cite{direct}. These include the effects of 
 nonuniversality of the scalar masses at the unification
 scale\cite{nonuni}, 
 effects of CP violation with EDM constraints\cite{cin},
 effects of coannihilation\cite{coanni}, the effects of the 
 $g_{\mu}-2$ constraint, as well as the effect of 
  variations of the
 WIMP velocity\cite{bottino1,roszkowski,corsetti1}, and the effects of
 rotation of the galaxy\cite{sikvie} in the prediction of detection rates
 for the direct and the indirect detection of dark matter. 
 In this work we focus on the effects of constraints of Yukawa
 unification on dark matter. This topic has largely not been
 addressed in the literature, except for the work of Ref.\cite{gomez}
 which, however, does not take account of gaugino mass nonuniversalities
 which is an important element of the present work.
 We focus on models where Yukawa unification occurs for $\mu$ positive
 (we use the sign convention of Ref.\cite{sugra}) consistent with
 the $b\rightarrow s+\gamma$ and the $g_{\mu}-2$ constraint from 
 Brookhaven.
 The outline of the rest of the paper is as follows: In Sec.2
 we discuss the framework of the analysis. 
 In Sec.3 we discuss the $g_{\mu}-2$ constraint which affects 
 dark matter analyses. This constraint requires a revision because  
 of a recent reevaluation of the light by light hadronic contribution
  to $g_{\mu}-2$. In Sec.4 we discuss 
 the results for the satisfaction of the relic density limits 
 under Yukawa unification constraint with $\mu>0$. In Sec.5 we discuss the 
 neutralino-proton cross sections including the effect of a new
 determination of the pion-nucleon sigma term. In Sec.6 we give conclusions.

\section{Theoretical Framework} 
  The primary quantity of interest in the study of dark matter
  is  $\Omega_{\chi} h^2$  where  
$\Omega_{\chi}=\rho_{\chi}/\rho_c$, where $\rho_{\chi}$ is the
neutralino relic density, $\rho_c=3H_0^2/8\pi G_N$ is the critical
matter density, and  $h$ is the value of the Hubble parameter $H_0$
in units of 100 km/sMpc. Experimentally the limit on $h$ from the
Hubble  Space Telescope is $h=0.71\pm 0.03\pm 0.07$\cite{freedman}.
The total $\Omega=\Omega_m +\Omega_{\Lambda}$ where $\Omega_m$
is the total matter density and $\Omega_{\Lambda}$ is the
dark energy density. For $\Omega_m$ we assume the simple model 
$\Omega_m=\Omega_B +\Omega_{\chi}$, where $\Omega_B$ is the baryonic
component and  $\Omega_{\chi}$ is the neutralino component which we assume
constitutes the entire dark matter. Using the 
 recent analysis of $\Omega_m$ which gives\cite{lineweaver} 
 $\Omega_m=0.3\pm 0.08$ and assuming $\Omega_B\simeq 0.05$, one finds 
 \beq
 \Omega_{\chi} h^2=0.126\pm 0.043
 \eeq
 With the above numerics and using a rather
liberal error corridor  we have  the following limits on 
$\Omega_{\chi} h^2$
 
 \begin{equation}
 0.02\leq \Omega_{\chi} h^2\leq 0.3
 \end{equation}
 In the determination of the neutralino relic density we  
 use the standard techniques and compute $\Omega_{\chi} h^2$
 using the formula
\beqn
\Omega_{\chi} h^2\cong 2.48\times 10^{-11}{\biggl (
{{T_{\chi}}\over {T_{\gamma}}}\biggr )^3} {\biggl ( {T_{\gamma}\over
2.73} \biggr)^3} {N_f^{1/2}\over J ( x_f )}\nonumber\\
J~ (x_f) = \int^{x_f}_0 dx ~ \langle~ \sigma \upsilon~ \rangle ~ (x) \rmGeV^{-2}
\eeqn
In the above  $\biggl({{T_{\chi}}\over {T_{\gamma}}}\biggr )^3$ is the
reheating factor,
 $N_f$ is the number of degrees of freedom at the freeze-out 
temperature $T_f$ and  $x_f= kT_f/m_{\tilde{\chi}}$.
The determination of $J(x_f)$ is carried out using the  accurate 
techniques  developed in Ref.\cite{accurate}.
 
 It is known that gaugino mass nonuniversalities can significantly
 affect neutralino relic densities and dark matter searches. 
 Specifically in Ref.\cite{corsetti2} an analysis of the effects of
 nonuniversalities of the gaugino masses on dark matter was carried
 out in the framework of SU(5) grand unification and in D brane models. 
 We note in passing that there is no rigid relationship between 
 the ratios of $SU(3)\times SU(2)\times U(1)$ gauge coupling constants
 at the GUT scale ($M_G \sim 2 \times 10^{16}$ GeV) and the 
ratio of $SU(3)$, 
 $SU(2)$,  and $U(1)$ gaugino masses
 at the GUT scale. The ratios of the gauge coupling constants 
 at the GUT scale are determined
 purely by the GUT group, while the gaugino masses  are soft 
 SUSY breaking parameters which involve both GUT and Planck scale physics.
  This topic has been discussed in several 
 works (see for example, Refs.\cite{corsetti2,anderson,eent} and 
 the references therein). For the present analysis we assume
 nonuniversality of gaugino masses and impose unification of gauge
 coupling constants at the GUT scale. 
 Returning to the general structure of the gaugino masses one finds
 that for the case of SU(5) the gaugino mass terms
 can arise from any of the representations that lie in the
 symmetric product of $24\times 24$. Since 
 
 \begin{equation}
 (24\times 24)_{sym}=1+24+75+200
 \end{equation}
 one finds that in general the gaugino masses are nonuniversal at the
 GUT scale arising from nonuniversalities due to the 24, 75 and 200 plet
 on the right hand side and one may in general write the 
 $SU(3)\times SU(2)\times U(1)$ gaugino masses as a sum  
 \begin{equation}
 \tilde m_i(M_G)=m_{\frac{1}{2}}\sum_r c_r n_i^r
\end{equation} 
 where  $n_i^r$ are characteristic of the representation $r$ and
 $c_r$ are relative weights of the representations in the sum.
 Specifically  
  the $SU(3)$,  $SU(2)$, and  $U(1)$ gaugino masses at  
  the GUT scale for different representations have the following 
  ratios\cite{anderson}: 
  $M_3:M_2:M_1$ are in the ratio $2:-3:-1$ for the 24 plet,
   in the ratio $1:3:-5$ for the 75 plet, and in the ratio $1:2:10$
  for the 200 plet. The 24 plet case is of special interest 
  for reasons which we now explain.  
  It is known that the sign of the supersymmetric
  contribution to $a_{\mu}$ is directly correlated with the sign of
 $\mu$\cite{kosower}
  in  mSUGRA\cite{msugra} and in other models which share the same generic
 features as mSUGRA. Thus in mSUGRA one finds that $a_{\mu}^{SUSY}>0$
 for $\mu>0$ and $a_{\mu}^{SUSY}<0$ for $\mu<0$.   
 Since experiment indicates $a_{\mu}^{SUSY}>0$ one infers that the sign of
 $\mu$ is positive\cite{chatto2}.
 The $\mu$ sign is also of relevance for the satisfaction of 
 the $b\rightarrow s+\gamma$ constraint. It is known that 
 the $b\rightarrow s+\gamma$ constraint favors a positive value of
 $\mu$\cite{bsgamma,bsgammanew}. However, a problem arises because
 $b-{\tau}$ unification appears to favor a negative
 value of $\mu$\cite{bagger,deboer}. 
 This is so because the supersymmetric correction
 to the b quark mass from the dominant gluino exchange 
 contribution\cite{susybtmass}
 is  negative for $\mu$ negative and a negative contribution to the
 b quark mass helps $b-\tau$ unification.
 Several works have appeared recently which explore ways to help
 resolve this problem\cite{bf,bdr,ky,cnbtau}.
  Specifically it was pointed out in Ref.\cite{cnbtau} that gaugino mass
  nonuniversalities possess a mechanism which can generate a negative 
  contribution to the b quark mass for a positive $\mu$.
  In the context of SU(5) the mechanism arises from the gaugino mass
  ratios coming from the 24 plet of SU(5) in Eq.(4).
  For this case it was shown that an $a_{\mu}^{SUSY} >0$
 naturally leads to a negative correction to the b quark mass  
  even for $\mu>0$. This phenomenon comes about because the gluino 
  exchange contribution to the b quark mass is proportional to 
  $\mu \tilde m_3$ and 
 the opposite  sign correlation between  $\tilde m_2$ and 
 $\tilde m_3$ natually leads to a negative contribution to the
 b quark mass. 
 In this case one
  finds that all the constraints including  $b-{\tau}$ unification and
  the  $b\rightarrow s+\gamma$ constraint are easily satisfied. 
 
 One can investigate the phenomenon discussed above also in the
 context of SO(10). Here in general the gaugino masses will lie
 in the symmetric SO(10) irreducible representations of $45\times 45$
 where 
 \begin{equation}
 (45\times 45)_{sym}=1+54+210+770
 \end{equation}
In this case the nonuniversalities of the gaugino masses arise
 due to the 54, 210 and  770 plets
 on the right hand side. Specifically, here one finds that 
 the 54 plet case can give rise to  patterns of 
 $SU(3)$, $SU(2)$, and $U(1)$ gaugino masses which look interesting 
 for $b-t-\tau$ unification. Thus for the symmetry breaking pattern
 $SO(10)\rightarrow SU(4)\times SU(2)\times SU(2)$
 $\rightarrow SU(3)\times S(2)\times U(1)$ one finds  that the
$SU(3)$, $SU(2)$, $U(1)$ gaugino masses at  the GUT scale are in 
the ratio $M_3:M_2:M_1 = 1:-3/2:-1$\cite{nonuniso10}.
  The $SU(3)$ and $SU(2)$ 
   gaugino masses here have  opposite signs and are similar to  
   the 24 plet case.  
Thus  this case looks favorable for generating negative
  corrections to the b quark mass consistent with $a_{\mu}^{SUSY}>0$
  and for the unification of
  Yukawa couplings for $\mu>0$.
  There is another pattern of symmetry breaking which also looks
  interesting. Thus the symmetry breaking pattern 
  $SO(10)\rightarrow SU(2)\times SO(7)$
 $\rightarrow SU(3)\times S(2)\times U(1)$ yields 
  $SU(3)$, $SU(2)$, and $U(1)$ gaugino masses at the GUT scale in the 
  ratio $M_3:M_2:M_1=1:-7/3:1$\cite{nonuniso10}. Here again  
  the $SU(3)$ and $SU(2)$ gaugino masses have opposite signs 
  and it appears possible to get negative corrections to the b quark
  mass necessary for $b-\tau$ unification consistent with 
  $a_{\mu}^{SUSY}>0$ and $\mu>0$.
  Thus we will discuss the following three cases in this paper:
  (a) 24 plet case: This is the model where the nonuniversalities
  originate from the 24 plet of SU(5) where $M_3:M_2:M_1=2:-3:-1$;
  (b) 54 plet case: This is the model where the 
  nonuniversalities originate from the 54 plet of SO(10) and the 
  symmetry breaking pattern gives $M_3:M_2:M_1=2:-3:-1$;
  (c) $54'$ plet case: This is the model  where the 
  nonuniversalities originate from the 54 plet of SO(10) and the 
  symmetry breaking pattern gives $M_3:M_2:M_1=2:-7/3:1$.
   The question that remains to be explored is
 what happens to supersymmetric dark matter in these models 
 in the parameter space
 which is consistent with Yukawa unification and consistent with other 
 constraints.  We will discuss this topic in Secs.4 and 5 after reviewing
 the revised constraint on $g_\mu-2$ in Sec.3.

\section{The Revised $g_{\mu}-2$ Constraint}
The recent Brookhaven experimental result gives\cite{brown}
$a_{\mu}^{exp}=11659203(15)\times 10^{-10}$ where  
$a_{\mu}=(g_{\mu}-2)/2$. The standard model prediction for this
quantity consists of\cite{czar1} the $O(\alpha^5)$ qed 
correction, the one and the two loop electro-weak corrections 
and the hadronic correction\cite{davier}. The hadronic correction
has been rather controversial\cite{hadronic}. It consists of
 the  $O(\alpha^2)$ and $O(\alpha^3)$ hadronic vacuum polarization,
 and the light-by-light hadronic contribution. 
 For the light by light hadronic contribution 
 two previous analyses gave
 the following values:
 $a_{\mu}^{LbyL;had}=-7.9(1.5)\times 10^{-10}$\cite{hayakawa}   
 and  $a_{\mu}^{LbyL;had}=-9.2(3.2)\times 10^{-10}$\cite{bijnens}.
 These give an average of $a_{\mu}^{LbyL;had}=-8.5(2.5)\times 10^{-10}$.
Using the  $O(\alpha^2)$ and $O(\alpha^3)$ hadronic vacuum polarization
analysis of Ref.\cite{davier} and the average light by light contribution 
as discussed above one finds 
$a_{\mu}^{exp}-a_{\mu}^{SM}=43(16)\times 10^{-10}$. 
However,  a very recent reevaluation of light by light contribution 
 finds \cite{knecht}
 $a_{\mu}^{LbyL;had}=+8.3(1.2)\times 10^{-10}$ which although 
 essentially of the same magnitude is opposite in sign to the 
 previous determinations. Spurred by the above, 
 Hayakawa and Kinoshita\cite{hkrevised} reexamined the
 light by light contribution and found  an error in sign in the treatment
  of the $\epsilon -$ tensor in the algebraic manipulation program FORM
   used in their analysis. Their revised value of  
 $a_{\mu}^{LbyL;had}=+8.9(1.54)\times 10^{-10}$ is now in good 
 agreement with the analysis of Ref.\cite{knecht}.  The average of the two
 evaluations gives  $a_{\mu}^{LbyL;had}=+8.6(1)\times 10^{-10}$.
 Correcting for the above one finds
 
 \begin{equation}
 a_{\mu}^{exp}-a_{\mu}^{SM}=26(16)\times 10^{-10}
 \end{equation}
 which is a $1.6\sigma$ deviation between experiment and
 theory. We discuss now the implications of this constraint relative 
 to the case when one had $a_{\mu}^{exp}-a_{\mu}^{SM}$ 
 $=43(16)\times 10^{-10}$. For the case when the $a_{\mu}^{exp}-a_{\mu}^{SM}$
 difference was taken to be $43(16)\times 10^{-10}$  one found 
 using a $2\sigma$ error
  corridor  interesting upper limits on
 the soft SUSY parameters and specifically for the mSUGRA case one 
 found that the upper limits on $m_0$ and $m_{\frac{1}{2}}$ were
 $m_0\leq 1.5$ TeV and $m_{\frac{1}{2}}\leq 800$ GeV for a range of
 $\tan\beta$ values of $\tan\beta\leq 55$\cite{chatto2}. 
 These ranges are well within
 the discovery limit of the Large Hadron Collider\cite{cms}.

 Since the reevaluated difference 
 $a_{\mu}^{exp}-a_{\mu}^{SM}$ is now less than $2\sigma$, we 
 consider a reduced error corridor to obtain meaningful constraints.
 We give here an analysis under two separate assumptions for the
 error corridor: one for $1.5\sigma$ and the other for $1\sigma$. 
 The results of the analysis with these error 
 corridors are exhibited in Figs.~\ref{amutan5}-~\ref{amutan55} 
for $\tan\beta$ values of
 5, 10, 30, 45 and 55 and $\mu>0$. The $1.5\sigma$ case  of Fig.~\ref{amutan5}
 with $\tan\beta =5$ gives the upper limits $m_0\leq 850$ GeV and 
  $m_{\frac{1}{2}}\leq 800$ GeV. However, here the lower 
limit of Higgs boson mass indicated
   by the LEP data lies outside of the allowed parameter space.
 The $1.5\sigma$ case of Fig.~\ref{amutan10}
 with $\tan\beta =10$ gives the upper limits $m_0\leq 1300$ GeV and 
  $m_{\frac{1}{2}}\leq 1100$ GeV. Here the parameter space includes the
  lower limit of the Higgs boson mass indicated by the LEP data.
    For the case 
  of Fig.~\ref{amutan30} with $\tan\beta =30$ one finds the upper limit of 
  $m_0\leq 2500$ GeV for the $1.5\sigma$ case which is on the
  borderline of  the reach  of the LHC\cite{cms} and most likely beyond
  its reach.  However, for the
  $1\sigma$ case one finds the upper limits  $m_0\leq 1000$ GeV and 
 $m_{\frac{1}{2}}\leq 800$ GeV which lie well within the discovery
 potential of the LHC. A similar situation holds for $\tan\beta=45$
 and for $\tan\beta=55$.
  For  the  $\tan\beta =45$ case of Fig.~\ref{amutan45} the upper limit for
 $m_0$ is $m_0\leq 2700$ GeV for the $1.5\sigma$ case while 
 the $1\sigma$ case gives 
 $m_0\leq 1300$ GeV and $m_{\frac{1}{2}}\leq 825$ GeV,
 which lie well within the reach of the LHC. Similarly the $\tan\beta=55$
 plot of  Fig.~\ref{amutan55} gives $m_0\leq 2500$ GeV for the $1.5\sigma$ case
 but the $1\sigma$ case
  gives $m_0\leq 1450$ GeV and $m_{\frac{1}{2}}\leq 625$ GeV.
  Again while the $1.5\sigma$ upper limits are on the border line
  of the reach of the LHC and most likely beyond its reach,
   the $1\sigma$ upper limits lie well within reach
  of the LHC. 
Since the upper limits for the $1.5\sigma$ case appear to cross 
the usual naturalness limits (see, e.g., Ref.\cite{ccn}) at least
for values of $\tan\beta \geq 30$ we impose    
 the constraint  of $1\sigma$ error corridor around the mean
 for $a_{\mu}^{exp} -a_{\mu}^{exp}$ as given by Eq.(7) for the
   analysis of Secs.4 an 5. The upper limits implied
   by the latter constraint lie well within the naturalness limits.

 \section{Yukawa Unification and Relic Density Analysis}
 We turn now to the main theme of the paper which is the analysis of
 dark matter under the Yukawa unification constraints. 
  As  discussed in  Ref.\cite{cnbtau} we define the Yukawa coupling unification
  parameter $\delta_{ij}$ for the Yukawa couplings $\lambda_i$ an
  $\lambda_j$ so that 
   \begin{equation}
  \delta_{ij} =\frac{|\lambda_i-\lambda_{j}|}{\lambda_{ij}}
  \end{equation} 
  where $\lambda_{ij}=(\lambda_{i}+\lambda_{j})/2$ and 
 $\delta_{ij}$ defines the degree of unification. 
  As is well known dark matter analysis is very sensitive to the
 $b\rightarrow s+\gamma$ constraint. 
 There are several recent  experimental determinations of 
 $b\rightarrow s+\gamma$ 
i.e., CLEO\cite{cleo}, BELLE\cite{belle} and \cite{aleph}
and we use their weighted  mean.
 Analyses of the theoretical prediction of the Standard Model
 including the leading order and the next to 
 leading order corrections for this branching ratio have been given
  by several authors\cite{gambino}. 
In our analysis we use a $2\sigma$ corridor in the difference 
between experiment and the prediction of the Standard Model  
to constrain our theoretical analysis of the supersymmetric contribution.  
 First we discuss the SU(5) case where we consider the gaugino
 mass nonuniversality at the GUT scale from the 24 plet representation
 as discussed in the paragraph following Eq.(5). The remaining soft 
 SUSY breaking parameters in the theory are assumed universal. 
 
 We begin by discussing the allowed parameter space in the 
 $m_0-C_{24}*m_{\frac{1}{2}}$ plane under the $g_{\mu}-2$ constraint.
 The results are exhibited in Fig.~\ref{gmu_24} for values of $\tan\beta$
 of 5, 10, 30 and 40. 
The top gray regions correspond to disallowed areas via radiative 
electroweak symmetry breaking requirement. 
The bottom patterned regions in Fig.~\ref{gmu_24} for the cases of 
$\tan\beta=$~5,10 and 30 are typically eliminated via the 
stability requirement of the Higgs potential at the GUT scale.
Part of the region 
with large $|c_{24} m_{1/2}|$ and large $m_0$ bordering the 
allowed (white) region for $\tanbeta=30$ is eliminated via the limitation of 
the CP-odd Higgs boson mass turning tachyonic at the tree level.  
For $\tanbeta=40$ most of the region 
(patterned and shaded) to the right of 
the allowed white small region is 
eliminated because of $\lambda_b$ going to 
the non-perturbative domain due to a large supersymmetric 
correction to the bottom mass.     
Regarding the SO(10) case, 
as is well known one needs to use nonuniversality of the Higgs boson 
masses at the GUT scale to achieve radiative breaking of the electroweak
  symmetry. For the analysis here we use 
 the nonuniversal Higgs scalar masses so that  
  $m_{H_1}^2=1.5m_0^2$ and $m_{H_2}^2=0.5m_0^2$. 
  The result of analysis of the allowed parameter space in 
the 54 plet case is given in Fig.~\ref{gmu_54} for values of 
$\tan\beta$ of 5, 10, 30 and 45.  The regions with patterns 
are discarded for reasons similar to as in Fig.~\ref{gmu_24}.
There is no discernible change in these results due to modest 
variations (up to 50\%) in the assumed nonuniversality 
(i.e., deviations of $m_{H_1}^2$ and $m_{H_2}^2$ at $M_G$ from $m_0^2$)  
of the soft parameters
 in the Higgs boson sector needed to accomplish radiative breaking
  of the electroweak symmetry.

We give now the relic density analysis.
 As a guide we  use the unification criterion $\delta_{b\tau}\leq 0.3$.
In Fig.~\ref{omega24a} we plot $\Omegachi h^2$ vs $\tan\beta$ for 
the following 
range of parameters: $0<m_0<2000$ GeV, 
$-1000 \rmGeV <c_{24}m_{\frac{1}{2}}<1000 \rmGeV$,
$-6000 \rmGeV <A_0<6000 \rmGeV$ and $\mu>0$. The small crosses are the 
points that satisfy the $g_{\mu}-2$ constraint, the filled squares 
additionally satisfy the 
$b\rightarrow s+\gamma$ constraint and the filled ovals  
satisfy all the constraints, ie., the $g_{\mu}-2$ constraint, 
the $b\rightarrow s+\gamma$ constraint and the $b-\tau$ unification 
constraint with $\delta_{b\tau}\leq 0.3$. One finds that there
exist significant regions of the parameter space as given by filled ovals 
where all the constraints are satisfied.  
The horizontal lines indicate the allowed corridor for the relic density
as given by Eq.(2). A plot of 
$\Omegachi h^2$ vs $m_0$ for exactly the same
ranges of the parameter space as in Fig.~\ref{omega24a} is given in Fig.~\ref{omega24b}.
A similar plot of $\Omegachi h^2$  as a function of 
$c_{24}*m_{\frac{1}{2}}$ is given in Fig.~\ref{omega24c}. 
In Fig.~\ref{omega24d} we give a plot of $\Omegachi h^2$ vs $A_0$ 
and in Fig.~\ref{omega24e} we give a plot of 
$\Omegachi h^2$ as a function of the LSP mass $m_{\chi}$. 
The paucity of points in the region around the neutralino mass
of 45 GeV in the allowed corridor of relic density in Fig.~\ref{omega24e} 
is due to the rapid s-channel Z pole annihilation and also due to the 
s-channel Higgs pole annihilation in the region below $m_{\chi}\sim 60$ GeV. 
Finally in Fig.~\ref{omega24f} we give a plot of $\Omegachi h^2$ vs $\delta_{b\tau}$.
One finds that there exist regions of the parameter space where 
${b-\tau}$ unification even at the level of a few percent consistent
with the relic density and other constraints can be satisfied.

 We discuss next the SO(10) case with gaugino mass nonuniversality
 of the type $M_3:M_2:M_1=1:-\frac{3}{2}:-1$  as given in the 
 paragraph following Eq.(6). As noted earlier in this case
 the pattern of relative signs
 of the gaugino masses is similar to that for the 24 plet case.
In the analysis we impose not only the $b-\tau$ Yukawa unification
 but also $b-t$ and $t-\tau$ Yukawa unification\cite{als}. 
 Extrapolating from the 
 SU(5) case we impose the following constraints on $\delta_{b\tau}$,
 $\delta_{bt}$ and $\delta_{t\tau}$: all  $\delta_{ij}\leq 0.3$.
  In Figs.~\ref{omega54a} we plot $\Omegachi h^2$ vs $\tan\beta$ for 
 for the same range of parameters as in Fig.~\ref{omega24a}. The symbols used 
 in Fig.~\ref{omega54a}, i.e., the 
 small crosses, the filled squares and the filled ovals also
 have the same meaning as in Fig.~\ref{omega24a} except that the filled ovals 
 now include all the Yukawa unification constraints, i.e., 
 $\delta_{b\tau}, \delta_{bt},\delta_{t\tau}\leq 0.3$.
   Fig.~\ref{omega54a} shows that  there
 exist significant regions of the parameter space as given by filled ovals where all the constraints including the Yukawa unification constraints
are satisfied.  The horizontal lines indicate the allowed corridor for
the relic density as given by Eq.(2). 
In Fig.~\ref{omega54b} we give a plot of 
$\Omegachi h^2$ vs $m_0$ for exactly the same
ranges of the parameter space as in Fig.~\ref{omega54a}.
A similar plot of $\Omegachi h^2$  as a function of 
$c_{54}*m_{\frac{1}{2}}$ is given in Fig.~\ref{omega54c}. 
In Fig.~\ref{omega54d} we give a plot of $\Omegachi h^2$ vs 
$A_0$ and in Fig.~\ref{omega54e} we give a plot of 
$\Omegachi h^2$ as a function of the LSP mass $m_{\chi}$. 
Again the paucity of points in the region around the neutralino mass
of 45 GeV in the allowed corridor of relic density in Fig.~\ref{omega54e} 
is due to the rapid Z pole annihilation. 
Finally in Fig.~\ref{omega54f} we give a plot of $\Omegachi h^2$ vs $\delta_{b\tau}$
and similar plots exist for $\Omegachi h^2$ vs $\delta_{bt}$ and
$\Omegachi h^2$ vs $\delta_{t\tau}$ but are not exhibited.
Interestingly in this case one finds  a  high density  
of points where the relic density constraint consistent with
 Yukawa unification at the level of a few percent is satisfied.
An analysis of the gaugino-higgsino content of the neutralino 
 over the parameter space of the model consistent with 
 $b-\tau$ unification and all the other constraints can be 
 gotten by examining the expansion of the LSP so that  
 \beq 
\chi=\alpha \tilde B+ \beta\tilde W_3 + \gamma\tilde H_1
+ \delta \tilde H_2
\eeq
For the 24 plet case one finds that typically over most of the 
parameter space $\alpha^2+\beta^2>0.75$ while for the 
54 plet case  over most of the parameter space one has 
$\alpha^2+\beta^2>0.03$. While the dominant component in all cases
is the Bino\cite{bino}, in the 24 plet case one could
also have a significant higgsino component to the LSP while for the 
54 plet case the Bino purity of the LSP is rather high. 
We have carried a similar analysis for the relic density for the
second SO(10) case discussed in Sec.2 where 
$M_3:M_2:M_1=1:-\frac{7}{3}:1$ (we  label it as the $54'$ case).  
We do not exhibit the details as
this case appears marginal in the sense that the allowed neutralino
mass range is very small after the LEP limit on $m_{\chi}>32.3$ GeV
is imposed. One should keep in mind, however, that the LEP limit 
is a generic limit and is not deduced specifically for the model under
 discussion. Still this case has rather low neutralino mass upper 
 limit in any case. For the sake of completeness we will discuss
 the neutralino-proton cross sections for this case also in Sec.5.

   The allowed mass ranges for the three cases discussed above, i.e.,
 $24$, $54$ and $54'$ cases  are given in 
Table 1. The spectrum of Table 1 satisfies
 all the desired constraints, i.e., $g_{\mu}-2$, $b\rightarrow s+\gamma$
 and Yukawa unification constraints as discussed above. 
 A remarkable aspect of Table 1 is that the $b-\tau$ unification
 constraint implies a rather low Higgs boson mass. Since $\tan\beta$ values
 for scenarios with $b-\tau$ unification imply rather high values
 of $\tan\beta$, the experimental lower limits from LEP for large 
 $\tan\beta$ are rather low, i.e., $m_h>91$ GeV\cite{lephiggs}. 
 Thus the higgs mass ranges listed in Table 1 are all consistent 
 with the current experimental limits. Further, these limits are
 tantalizingly close to observation at RUNII of the Tevatron.
 Further, the spectrum for all the three cases listed Table I 
 is accessible to the LHC. As noted earlier the $54'$ case has only a very 
 narrow allowed range in neutralino mass and could be tested 
 or eliminated by data with a modest improvement in energy.

 
\begin{center}
\begin{tabular}{|c|c|c|c|}
\multicolumn{4}{c}{Table 1: ~Sparticle mass ranges for 24, 54, and $54'$ cases } \\
\hline
  Particle & {\bf 24} (GeV) & {\bf 54} (GeV) & {\bf $54'$} (GeV)\\
\hline
  $\chi_1^0$ & 32.3 - 75.2 & 32.3 - 81.0 & 32.3 - 33.4\\
 \hline
  $\chi_2^0$ & 96.7 - 422.5 & 94.7 - 240.8 & 145.7 - 153.9\\
 \hline
  $\chi_3^0$ & 110.5 - 564.3 & 301.5 - 757.1 & 420.9 - 633.8\\
 \hline
  $\chi_4^0$ & 259.2 - 575.9 & 311.5 - 759.7 & 427.6 - 636.9\\
 \hline
  $\chi_1^\pm$ & 86.9 - 422.6 & 94.6 - 240.8 & 145.8 - 153.9\\
 \hline
  $\chi_2^\pm$ & 259.9 - 577.2 & 315.1 - 761.6 & 430.7 - 639.2\\
 \hline
  $\tilde g$ & 479.5 - 1077.2 & 232.5 - 580.3 & 229.8 - 237.4 \\
 \hline
  $\tilde\mu_1$ & 299.7 - 1295.9 & 480.5 - 1536.8 & 813.1 - 1196.3\\
 \hline
  $\tilde\mu_2$ & 355.1 - 1309.3 & 489.8 - 1482.7 & 835.3 - 1237.6\\
 \hline
  $\tilde\tau_1$ & 203.5 - 1045.1 & 294.2 - 1172.6 & 579.4 - 863.7\\
 \hline
  $\tilde\tau_2$ & 349.6 - 1180.9 & 422.6 - 1311.7 & 704.6 - 1018.3\\
 \hline
  $\tilde u_1$ & 533.6 - 1407.2 & 566.7 - 1506.4 & 822.9 - 1199.8\\
 \hline
  $\tilde u_2$ & 561.1 - 1443.0 & 584.7 - 1544.6 & 849.6 - 1232.6\\
 \hline
  $\tilde d_1$ & 535.1 - 1407.5 & 580.3 - 1546.2 & 845.1 - 1232.5\\
 \hline
  $\tilde d_2$ & 566.7 - 1445.2 & 590.1 - 1546.7 & 853.3 - 1235.2\\
 \hline
  $\tilde t_1$ & 369.9 - 975.2 & 271.5 - 999.6 & 513.7 - 819.9\\
 \hline
  $\tilde t_2$ & 513.7 - 1167.6 & 429.4 - 1107.4 & 599.4 - 848.2\\
 \hline
  $\tilde b_1$ & 488.2 - 1152.8 & 158.1 - 1042.0 & 453.2 - 749.9\\
 \hline
  $\tilde b_2$ & 532.3 - 1207.0 & 396.6 - 1159.2 & 610.5 - 880.4\\
 \hline
  $h$ & 104.3 - 114.3 & 103.8 - 113.3 & 108.1 - 110.9\\
 \hline
  $H$ & 111.9 - 798.8 & 151.5 - 1227.6 & 473.4 - 831.9\\
 \hline
  $A$ & 110.5 - 798.8 & 151.4 - 1227.6 & 473.4 - 831.9\\
 \hline
  $\mu$ & 96.0 - 559.5 & 291.1 - 752.7 & 413.1 - 628.4\\
 \hline
 \end{tabular}\\
\end{center}

\section{Maximum and minimum neutralino-proton cross section}
In the analysis of neutralino-proton cross section $\sigma_{\chi -p}$
we restrict ourselves to the CP conserving case. Here the 
$\chi -p$ scattering is governed by the four Fermi interaction
${\cal L}_{eff}=\bar{\chi}\gamma_{\mu} \gamma_5 \chi \bar{q}
\gamma^{\mu} (A P_L +B P_R) q
+ C\bar{\chi}\chi  m_q \bar{q} q
+D  \bar{\chi}\gamma_5\chi  m_q \bar{q}\gamma_5 q
+E\bar\chi i \gamma_5 \chi m_q \bar q q
+F\bar \chi \chi m_q \bar q i \gamma_5 q$.  
We are specifically interested in neutralino scattering from heavy targets.
This scattering is dominated   
by the scalar interactions and in this case the $\chi -p$  cross-section
is given by 
\begin{equation}
 \sigma_{\chi p}(scalar)=\frac{4\mu_r^2}{\pi}
 (\sum_{i=u,d,s}f_i^pC_i+\frac{2}{27}(1-\sum_{i=u,d,c}f_i^p)
 \sum_{a=c,b,t}C_a)^2
 \end{equation}
In the above $f_i^p$ are the (i=u,d,s) quark densities
which are  defined by
 $m_pf_i^p=<p|m_{qi}\bar q_iq_i|p>$, and $\mu_r$ is the
 reduced mass. The scalar interaction parametrized by C arises 
 from several sources: from s channel contributions from the higgs 
 $h^0, H^0$ exchanges and from  t channel contributions from the sfermion
 exchange so that $C=C_{h^0}+C_{H^0}+C_{\tilde{f}}$. It was shown in
 Ref.\cite{corsetti2} that it is convenient to parameterize 
 the form factors $f_i^{(p,n)}$ such that  
 \beqn
f_{(u,d)}^p=m_{(u,d)}(m_u+m_d)^{-1}(1\pm\xi)\sigma_{\pi N}m_p^{-1},~~
f_s^p=m_s(m_u+m_d)^{-1}(1-x)\sigma_{\pi N}m_p^{-1}\nonumber\\
f_{(u,d)}^n=m_{(u,d)}(m_u+m_d)^{-1}(1\mp\xi)\sigma_{\pi N}m_p^{-1},~~
f_s^n=m_s(m_u+m_d)^{-1}(1-x)\sigma_{\pi N}m_p^{-1}
\eeqn
where $\sigma_{\pi N}$, x and  $\xi$ are defined by
$\sigma_{\pi N}$=$<p|2^{-1}(m_u+m_d)(\bar uu+\bar dd|p>$,
 $\xi=$$<p|\bar uu-\bar dd|p>(<p|\bar uu+\bar  dd|p>)^{-1}$, and
$x=\sigma_0/\sigma_{\pi N}$$=<p|\bar uu+\bar dd-2\bar ss|p>$
$(<p|\bar uu+\bar  dd|p>)^{-1}$.
Quark densities for the neutron are
related to the proton quark densities 
by\cite{corsetti2}$f_u^pf_d^p=f_u^nf_d^n$. 
Baryon mass splittings can be used to determine 
 the ratio $\xi/x$ and one finds\cite{corsetti2} $\xi/x=0.196$.
Using various determinations of $\sigma_0$ and $\sigma_{\pi N}$,
$x$ was estimated in Ref.\cite{corsetti2} to be $x=0.67\pm 0.18$ which
gives\cite{corsetti2} $\xi=0.132\pm 0.035$. 
Using the current data on the quark masses  one finds
$f_u^p=0.021\pm 0.004$,
$f_d^p=0.029\pm 0.006$, and
$f_s^p=0.21\pm 0.12$ and  
$f_u^n=0.016\pm 0.003$,
$f_d^n=0.037\pm 0.007$, and
$f_s^n=0.21\pm 0.12$.

It has been pointed out recently\cite{Bottino:2001} that an 
analysis of $\sigma_{\pi N}$\cite{Pavan:2001wz} using
new pion-nucleon scattering data\cite{said} leads to a significantly
larger neutralino-nucleon cross section. Thus the new determination
of  $\sigma_{\pi N}$\cite{Pavan:2001wz} lies in the range
$ 55 ~MeV \leq \sigma_{\pi N}\leq 73 ~MeV$ which is much larger than
the previous determinations (see, e.g., Ref.\cite{corsetti2}).
Using the new determination of $\sigma_{\pi N}$ and repeating 
the analysis of Ref.\cite{corsetti2} we find 
$x=0.55\pm 0.12$ and $\xi =0.108 \pm 0.024$. These lead to the
following new  determinations for the  quark 
densities
\beqn
f_u^p=0.027\pm 0.005,~
f_d^p=0.038\pm 0.006,~
f_s^p=0.37\pm 0.11\nonumber\\
f_u^n=0.022\pm 0.004,~
f_d^n=0.049\pm 0.007,~
f_s^n=0.37\pm 0.11
\eeqn
We use these new quark densities in our numerical analysis.
In Fig.~\ref{sigma24mchi} we exhibit the neutralino-proton
scalar cross section vs the neutralino mass $m_{\chi}$ for the
nonuniversal case of Fig.~\ref{omega24}.
The gaps in Fig.~\ref{sigma24mchi} are due to the relic density constraint as
can be seen from Fig.~\ref{omega24c} and Fig.~\ref{omega24e}.
 The DAMA 
region\cite{dama},
the lower limit achieved by CDMS\cite{cdms} and the future lower 
limits that may 
be achieved\cite{genius,cline} are also exhibited. First,
one finds that the allowed  
neutralino range is significantly reduced in this scenario with
mass range limited to less than 65 GeV. Second, one finds that
 the parameter space of the model can be fully probed 
by the proposed future dark matter detectors\cite{genius,cline}.
In this model  $\sigma_{\chi p}(scalar)$ lies in the range
\begin{equation}
 4\times 10^{-45}~(cm)^2\leq  \sigma_{\chi p}(scalar)\leq
 4\times 10^{-41}~(cm)^2  
\end{equation}
 In Fig.~\ref{sigma24tan}
we give a plot of $\sigma_{\chi p}(scalar)$ vs $\tan\beta$ which shows
that the upper limits of $\sigma_{\chi p}(scalar)$ are strongly
dependent on $\tan\beta$ as expected. 
An analysis of $\sigma_{\chi p}(scalar)$ vs $m_{\chi}$ for the 
nonuniversal SO(10) gaugino mass case of Fig.~\ref{omega54c} is given in
Fig.~\ref{sigma54mchi}.  Here the neutralino mass range extends up to 80 GeV.
As for the case of Fig.~\ref{sigma24mchi} the gaps in 
Fig.~\ref{sigma54mchi} are due to the
relic density constraint as can be seen from Fig.~\ref{omega24c} 
and Fig.~\ref{omega24e}.
Again as in the 24 plet case  all of the parameter space of this
 model can be fully probed by
the proposed future dark matter detectors\cite{genius,cline}.
In this model  $\sigma_{\chi p}(scalar)$ lies in the range\
\begin{equation}
 7\times 10^{-45}~(cm)^2\leq  \sigma_{\chi p}(scalar)\leq
 1\times 10^{-41}~(cm)^2
\end{equation}
In Fig.~\ref{sigma54tan} we give a plot of 
$\sigma_{\chi p}(scalar)$ vs $\tan\beta$. Here since $\tan\beta$ does
not vary over a wide range one does not see a  large enhancement
of $\sigma_{\chi p}(scalar)$ with $\tan\beta$ in this limited range. 
Finally, we discuss $\sigma_{\chi p}(scalar)$ for the $54'$ case.
In Fig.~\ref{sigma54primemchi} 
we exhibit $\sigma_{\chi p}(scalar)$ as a function of
$m_{\chi}$. In Fig.~\ref{sigma54primetan} 
we give a plot of $\sigma_{\chi p}(scalar)$ 
as a function of $\tan\beta$.  The lower and upper limits  on the 
scalar cross section in this case are very similar to the 54 plet 
case of Figs.~\ref{sigma54mchi} and ~\ref{sigma54tan}. However, 
as discussed in Sec.4 imposition
 of the lower limit of $32.3$ GeV on the neutralino mass eliminates
 most of the parameter space of this model.

\section{Conclusions}
In this paper we have given an analysis of supersymmetric 
dark matter under the constraint of Yukawa coupling 
unification with $\mu>0$.
The constraints of $b\rightarrow s+\gamma$ and the revised $g_{\mu}-2$
constraint taking account of the recent reevaluation of the light by
light hadronic correction were also imposed. The analysis was done 
exploiting the recently discovered region of the parameter space which 
utilizes 
nonuniversal gaugino masses and leads to negative corrections
to the b quark mass necessary for Yukawa coupling unification
with $\mu>0$. We considered scenarios
with  SU(5) and SO(10) unifications. Within SU(5) we considered 
nonuniversalities arising from the 24 plet representation of SU(5)
which allow for significant regions of the parameter space consistent
with $b-\tau$ unification so that $\delta_{b\tau}\leq 0.3$ for $\mu>0$
consistent with other constraints. This scenario limits the 
neutralino mass range to lie below 
65 GeV and within this range a significant part of the parameter space
 is consistent with
the relic density constraint. An analysis of the neutralino-proton
scalar cross section reveals that the allowed range of cross sections
can be fully probed by the proposed future dark matter detectors.
Within SO(10) we considered 
nonuniversalities arising from the 54 plet representation of SO(10)
which allow for significant regions of the parameter space consistent
with $b\tau, bt$ and $t\tau$ unification  constraints such that  
$\delta_{b\tau},\delta_{bt}, \delta_{t\tau}\leq 0.3$ for $\mu>0$
consistent with other constraints.  
 In this case one finds 
that the neutralino mass range extends till 80 GeV and again
the analysis of neutralino-proton
scalar cross section shows that the allowed range of cross sections
can be fully probed by the proposed future dark matter detectors.
One of the important features of models with $b-\tau$ unification 
explored here is a relatively low lying light Higgs boson with 
mass lying below 115 GeV. This mass range would certainly be
explored by RUNII of the Tevatron. Further, the entire sparticle
spectrum predicted in the class of models with $b-\tau$ unification
discussed here  would be accessible at the LHC. 
It would be interesting to explore SUSY signals such as the trileptonic
signal\cite{trilep} at colliders from this model. But such an 
investigation is beyond the scope of this paper.
	   
\noindent
\section{Acknowledgments}
Part of this work was done when P.N. was at CERN
and U.C.was visiting CERN and the Abdus Salam International
Center for Theoretical Physics, Trieste. They acknowledge 
the hospitality accorded to them. Computational facilities of 
ASCC at Northeastern University are acknowledged.
This research was supported in part by NSF grant PHY-9901057.

\begin{figure}           
\vspace*{-1.0in}                                 
\subfigure[]{                       
\label{amutan5} 
\hspace*{-0.6in}                     
\begin{minipage}[b]{0.5\textwidth}                       
\centering
\includegraphics[width=\textwidth,height=0.65\textwidth]{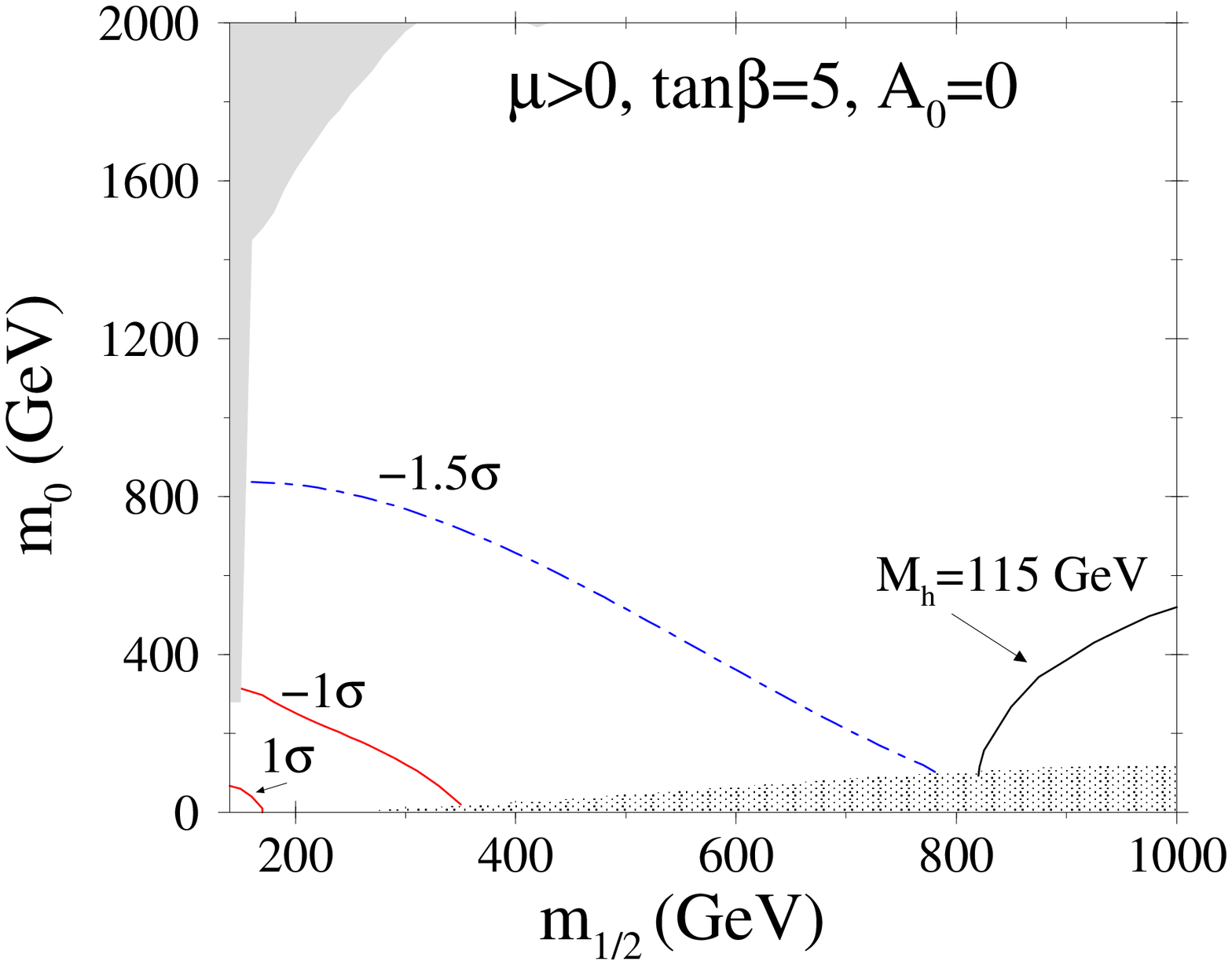}    
\end{minipage}}                       
\hspace*{0.3in}
\subfigure[]{      
\label{amutan10}                  
\begin{minipage}[b]{0.5\textwidth}                       
\centering                      
\includegraphics[width=\textwidth,height=0.65\textwidth]{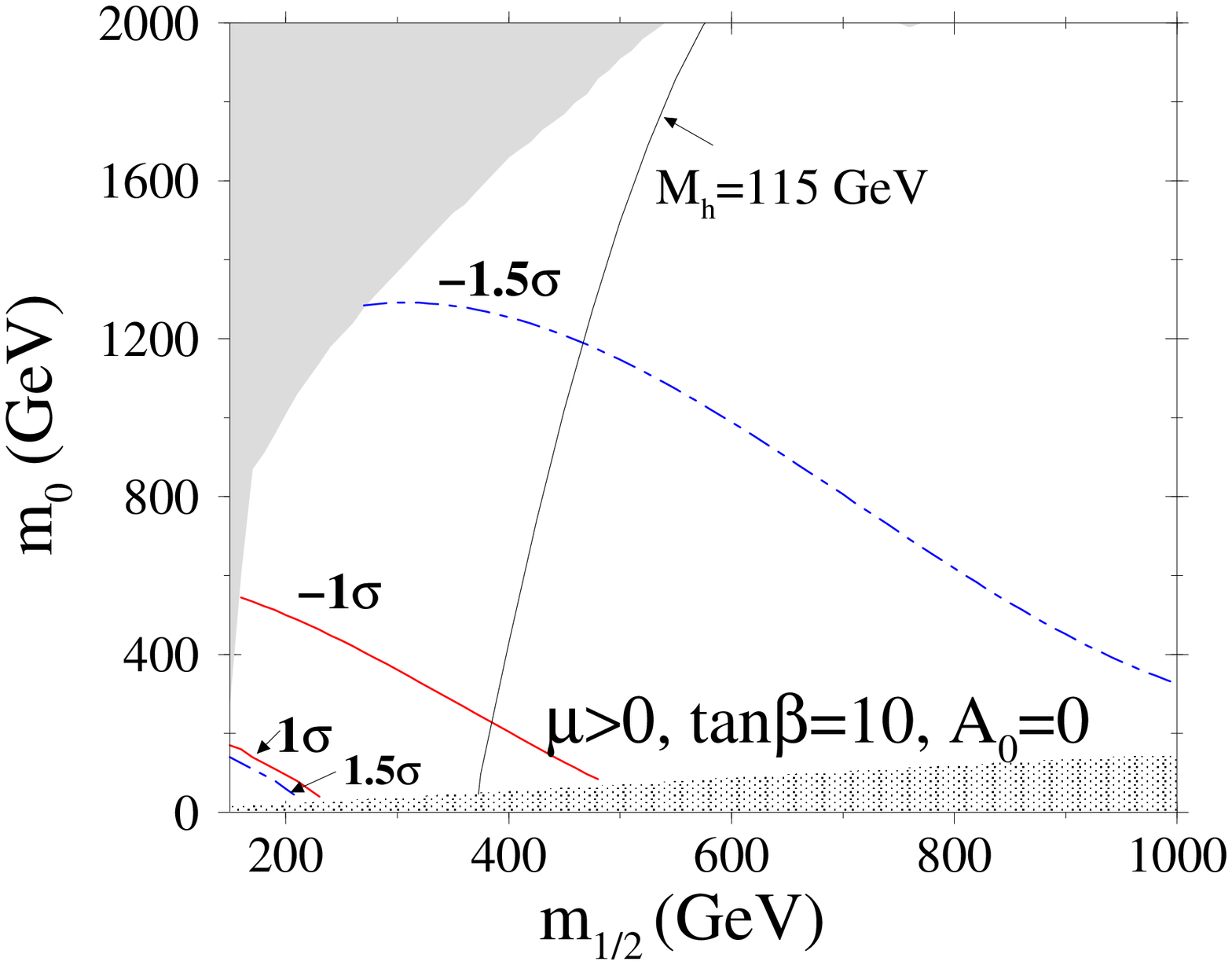} 
\end{minipage}}                       
\hspace*{-0.6in}                     
\subfigure[]{                       
\label{amutan30}                  
\begin{minipage}[b]{0.5\textwidth}                       
\centering
\includegraphics[width=\textwidth,height=0.65\textwidth]{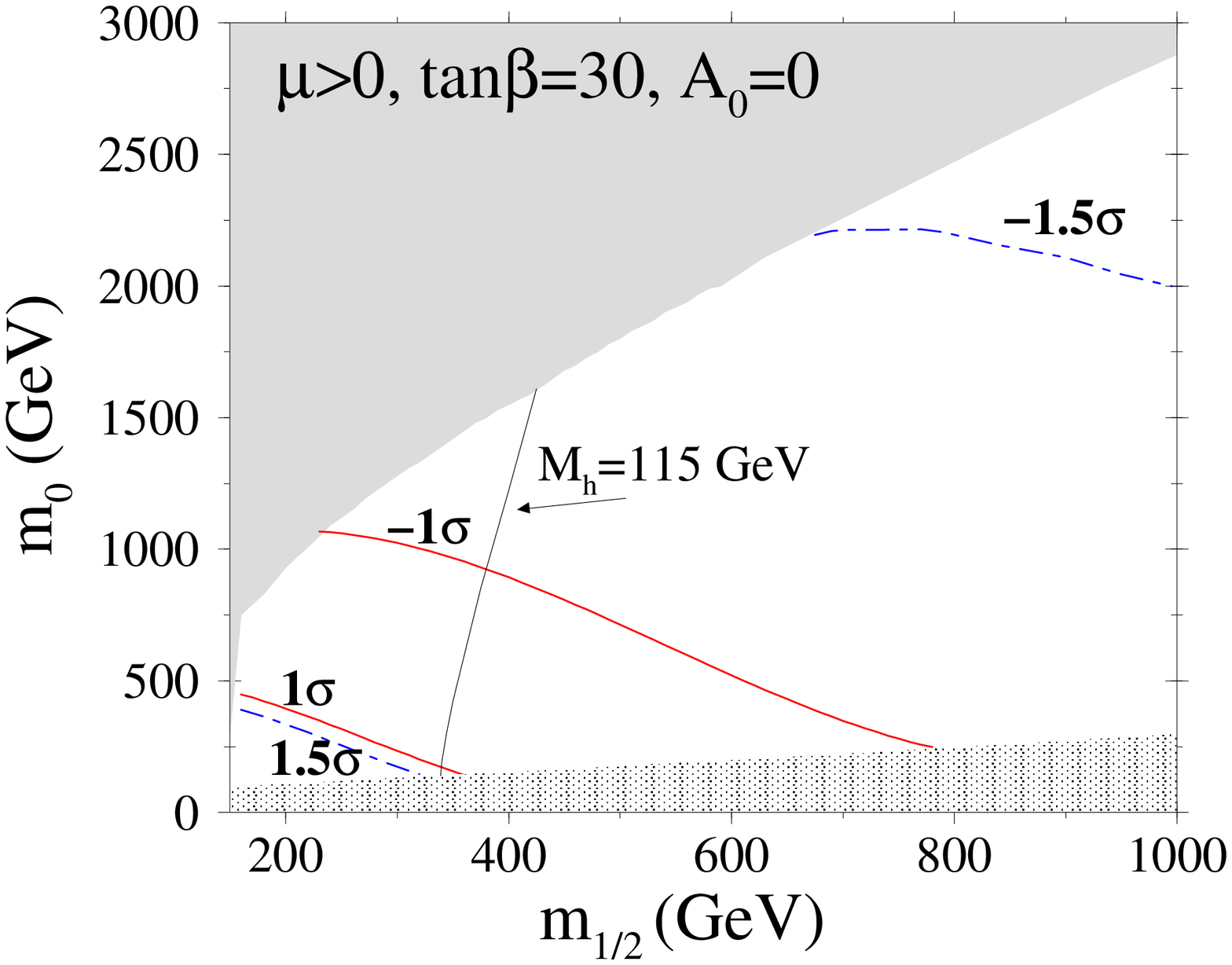}
\end{minipage}}
\hspace*{0.3in}                       
\subfigure[]{                       
\label{amutan45}
\begin{minipage}[b]{0.5\textwidth}                       
\centering                      
\includegraphics[width=\textwidth,height=0.65\textwidth]{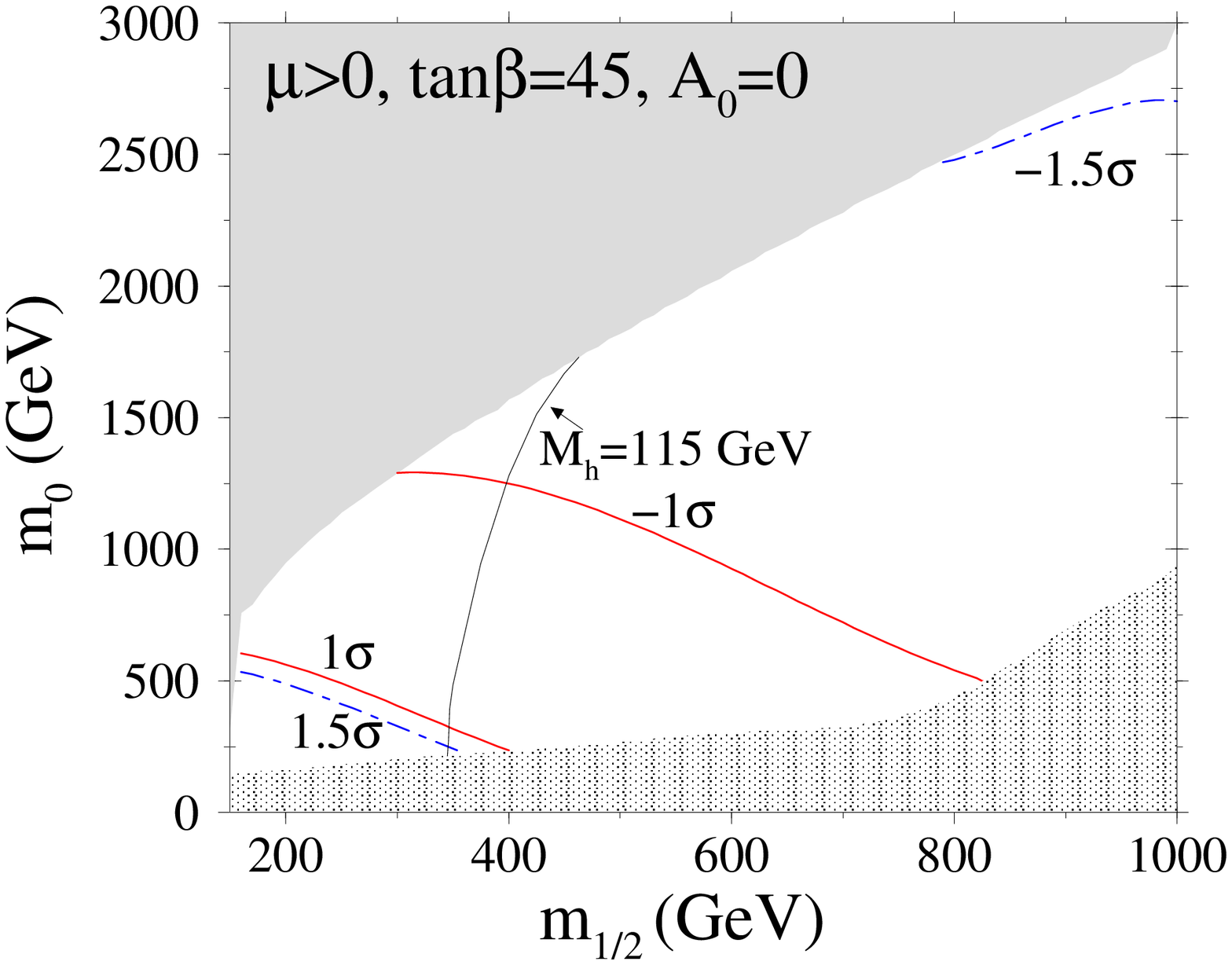}
\end{minipage}}
\begin{center}
\subfigure[]{                       
\label{amutan55}
\begin{minipage}[b]{0.5\textwidth}                       
\centering                      
\includegraphics[width=\textwidth,height=0.65\textwidth]{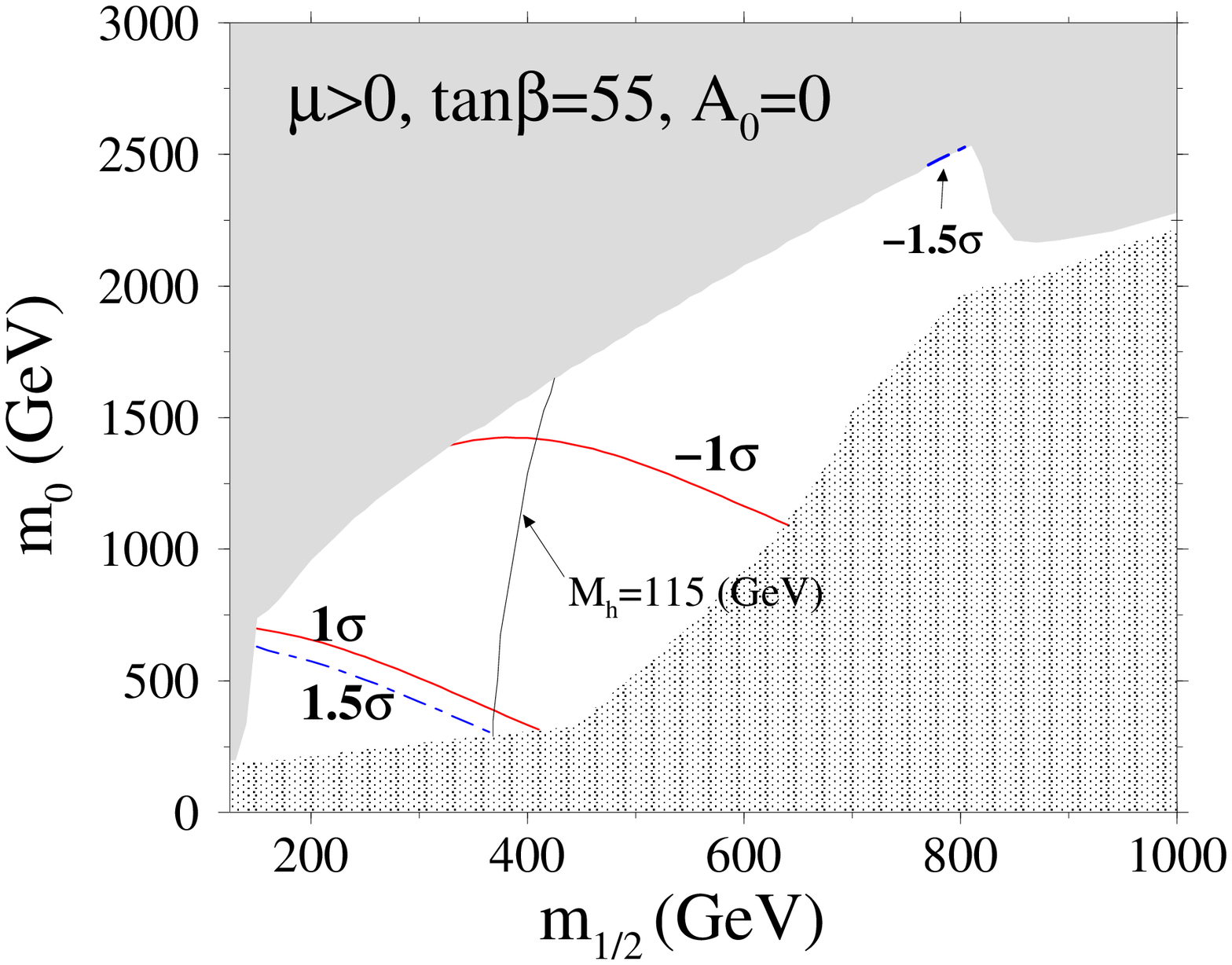}
\end{minipage}}
\end{center}                       
\caption{}                       
\label{amutan} 
\end{figure}

\newpage
\noindent
{\bf Figure Captions}\\
Fig.~\ref{amutan}:\\
Fig.~\ref{amutan5}:  
Allowed $a_{\mu}^{SUSY}$ regions corresponding to \
 the $1.5\sigma$ and the 
$1\sigma$  constraints for $\tan\beta=5$.  Similar analyses for
 $\tan\beta =10,30,45, 55$ are given in Fig.~\ref{amutan10}, 
Fig.~\ref{amutan30}, Fig.~\ref{amutan45}
 and Fig.~\ref{amutan55} respectively.  The top left gray regions do not 
satisfy the radiative electroweak symmetry breaking requirement whereas
 the bottom 
patterned regions are typically discarded by stau becoming the LSP. For 
large $\tan\beta$, 45 or 55, the bottom patterned region near the 
higher $m_{1/2}$ side and on the border of the white allowed regions are 
discarded because of CP-odd Higgs boson turning tachyonic at the tree level.\\

\begin{figure}           
\vspace*{-1.0in}                                 
\subfigure[]{                       
\label{gmu_24} 
\hspace*{-0.6in}                     
\begin{minipage}[b]{\textwidth}                       
\centering
\includegraphics[width=\textwidth,height=0.60\textwidth]{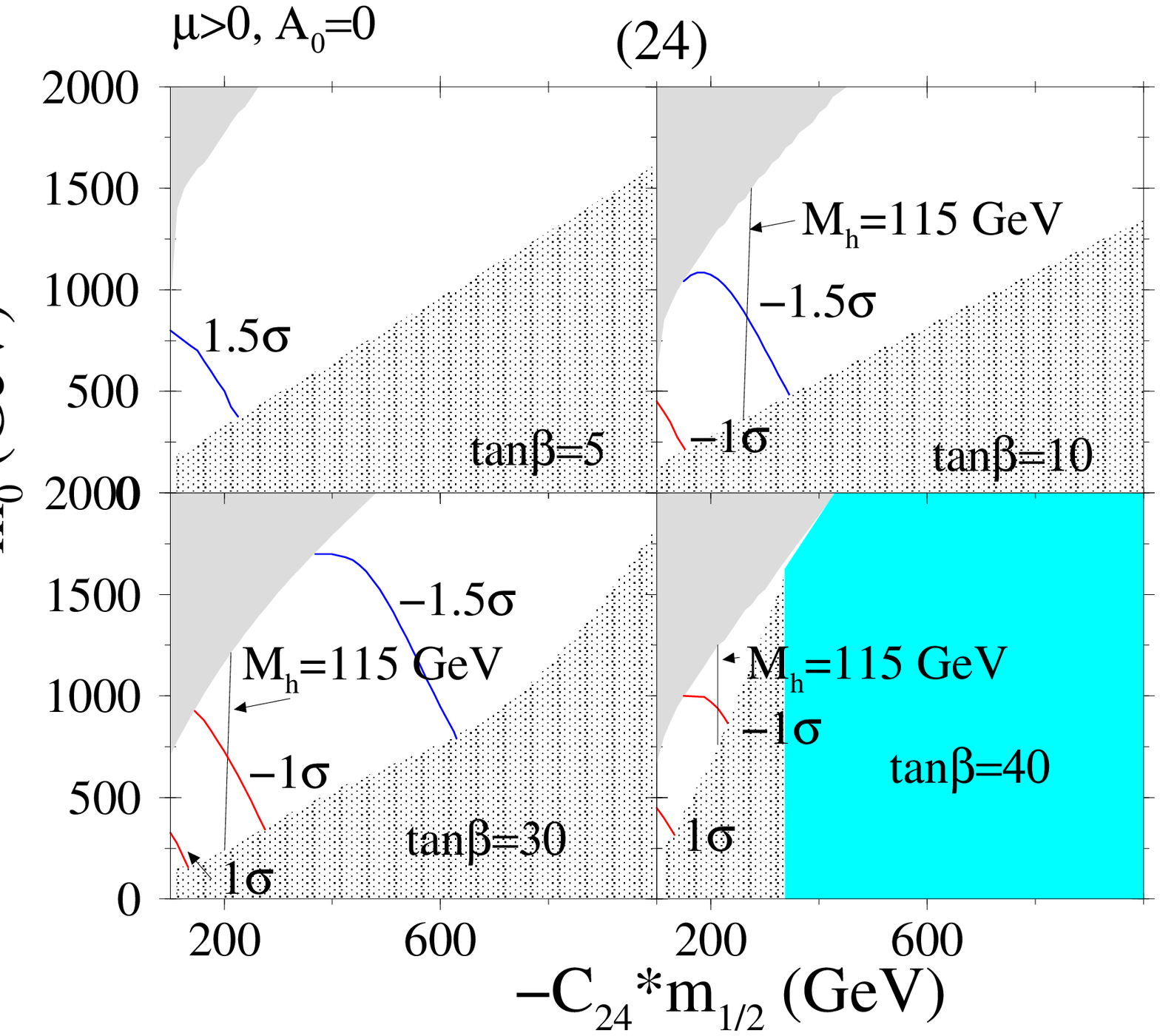}
\end{minipage}}

\subfigure[]{    
\label{gmu_54}                    
\hspace*{-0.6in}                      
\begin{minipage}[b]{\textwidth}                       
\centering
\includegraphics[width=\textwidth,height=0.60\textwidth]{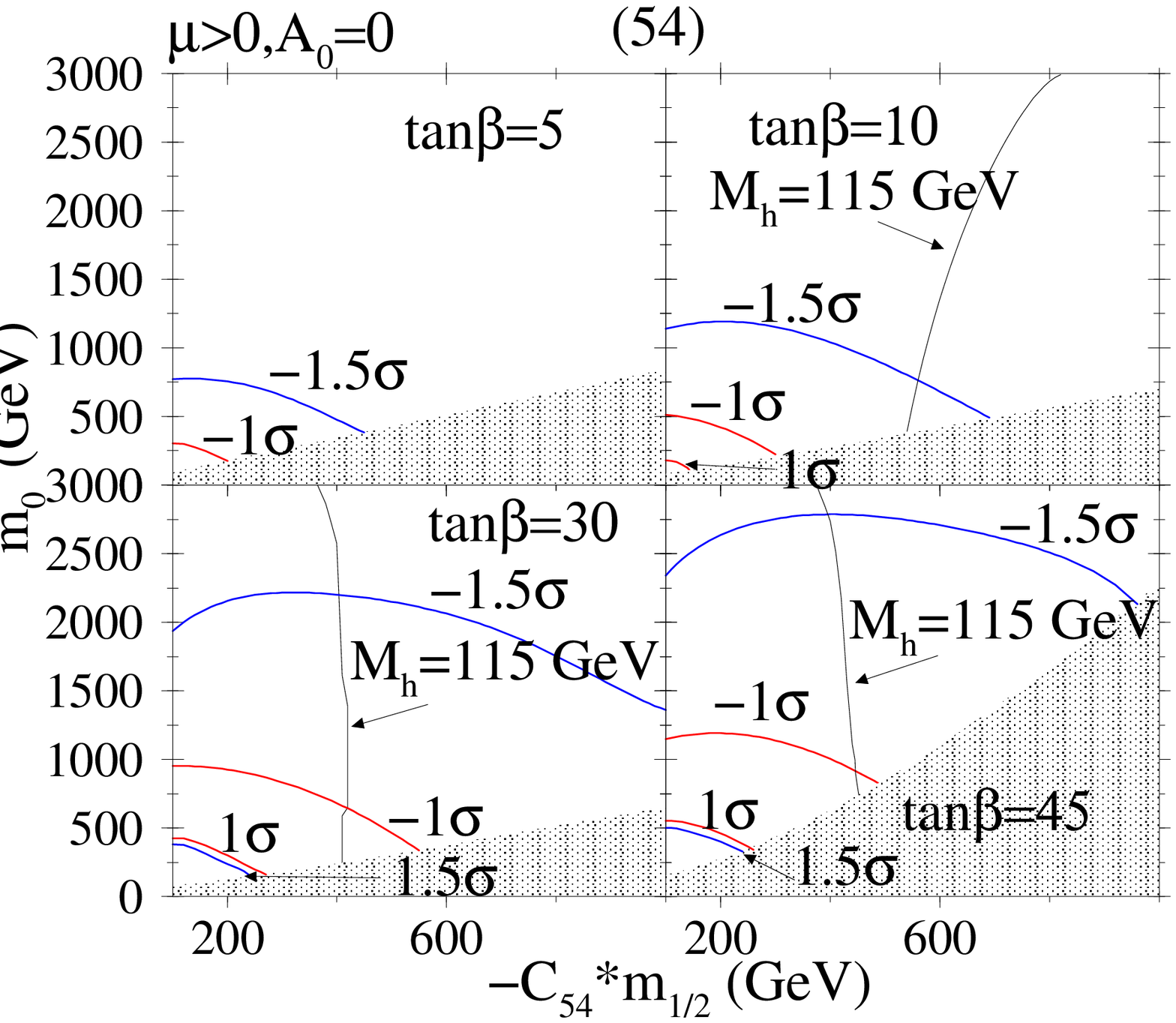}
\end{minipage}}
\caption{}                       
\label{gmu_nonuniv}
\end{figure}

\noindent
Fig.~\ref{gmu_nonuniv}:\\
Fig.~\ref{gmu_24}:
Allowed $g_\mu-2$ regions  corresponding to  $1.5\sigma$ and  
$1\sigma$  constraints for nonuniversal gaugino mass scenario of 
the SU(5) 24 plet case. A discussion of the discarded regions in 
the top and the bottom parts of the figures is given in the
text in Sec.4.\\
Fig.~\ref{gmu_54}:
Allowed $g_\mu-2$ regions corresponding to $1.5\sigma$ and 
$1\sigma$ constraints for nonuniversal gaugino mass scenario of 
SU(10) 54 plet case. Here the nonuniversal Higgs scalar 
parameters are given by $m_{H_1}^2=1.5m_0^2$ and $m_{H_2}^2=0.5m_0^2$.
A discussion of the discarded regions in 
the top and the bottom parts of the figures in given in Sec.4. \\

\begin{figure}           
\vspace*{-2.0in}                                 
\subfigure[]{                       
\label{omega24a} 
\hspace*{-0.6in}                     
\begin{minipage}[b]{0.5\textwidth}                       
\centering
\includegraphics[width=\textwidth,height=0.65\textwidth]{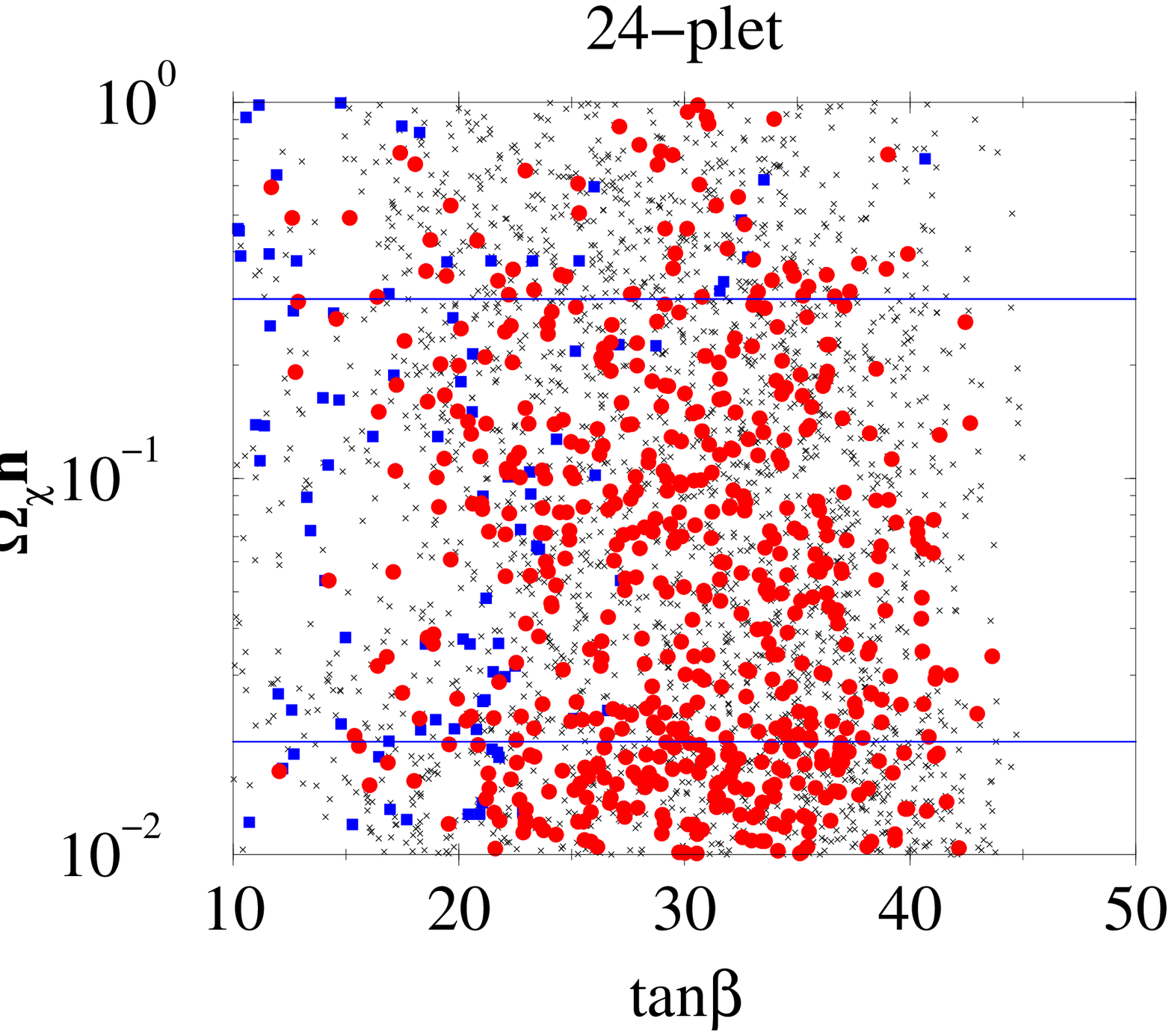}
\end{minipage}}                       
\hspace*{0.3in}
\subfigure[]{    
\label{omega24b}   
\begin{minipage}[b]{0.5\textwidth}                       
\centering                      
\includegraphics[width=\textwidth,height=0.65\textwidth]{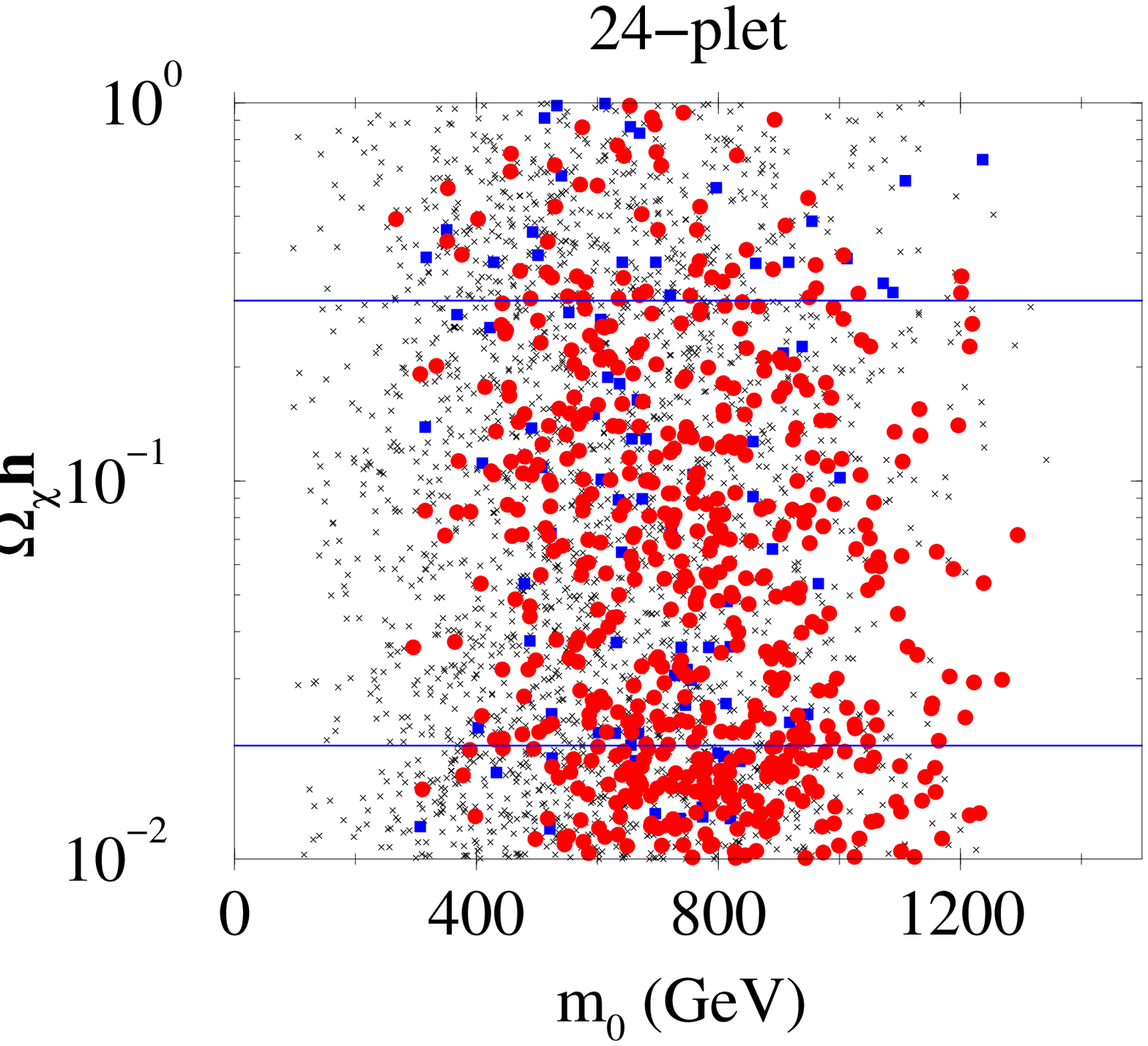} 
\end{minipage}}                       
\hspace*{-0.6in}                     
\subfigure[]{    
\label{omega24c}                    
\begin{minipage}[b]{0.5\textwidth}                       
\centering
\includegraphics[width=\textwidth,height=0.65\textwidth]{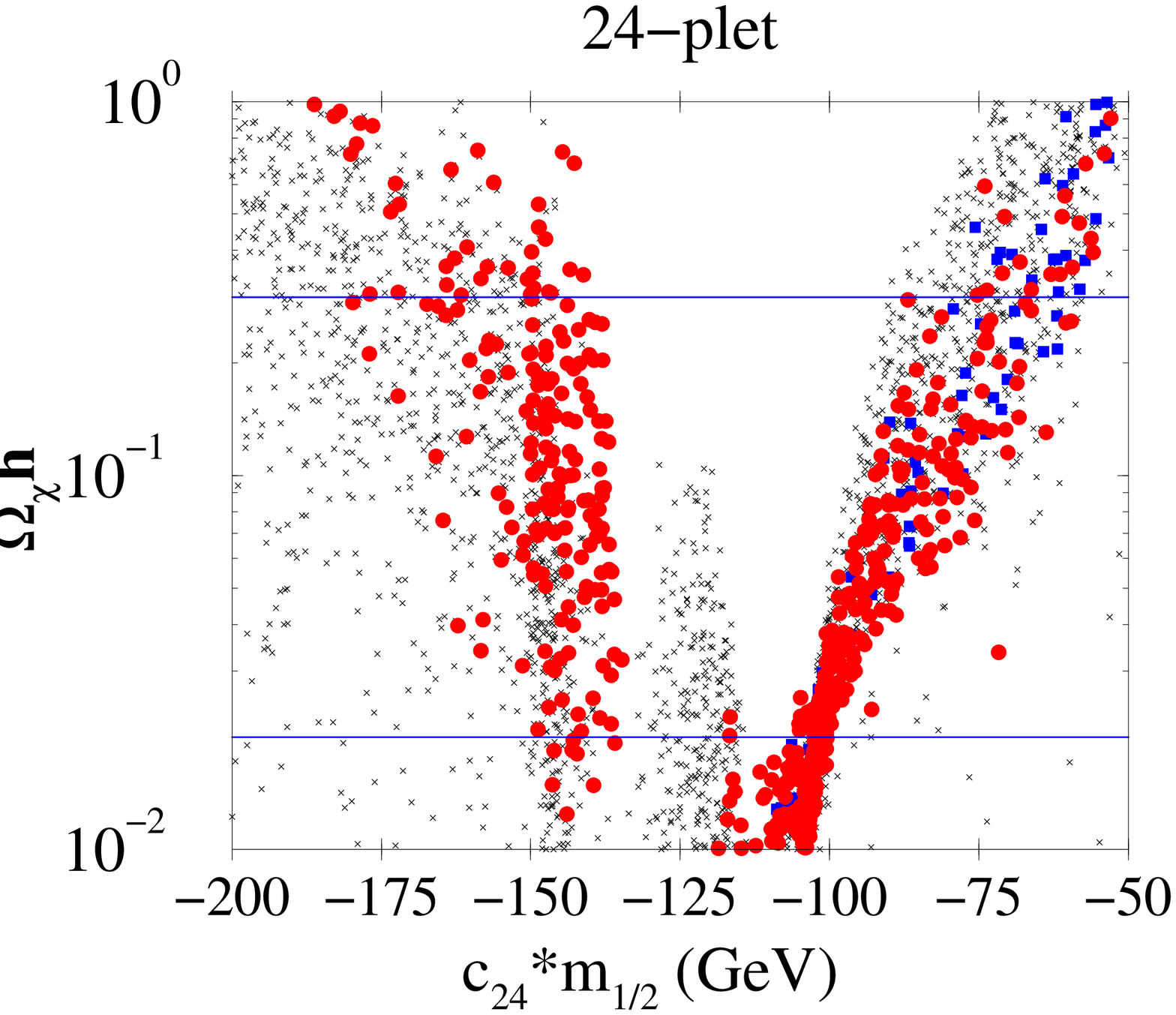}
\end{minipage}}
\hspace*{0.3in}                       
\subfigure[]{    
\label{omega24d}                    
\begin{minipage}[b]{0.5\textwidth}                       
\centering                      
\includegraphics[width=\textwidth,height=0.65\textwidth]{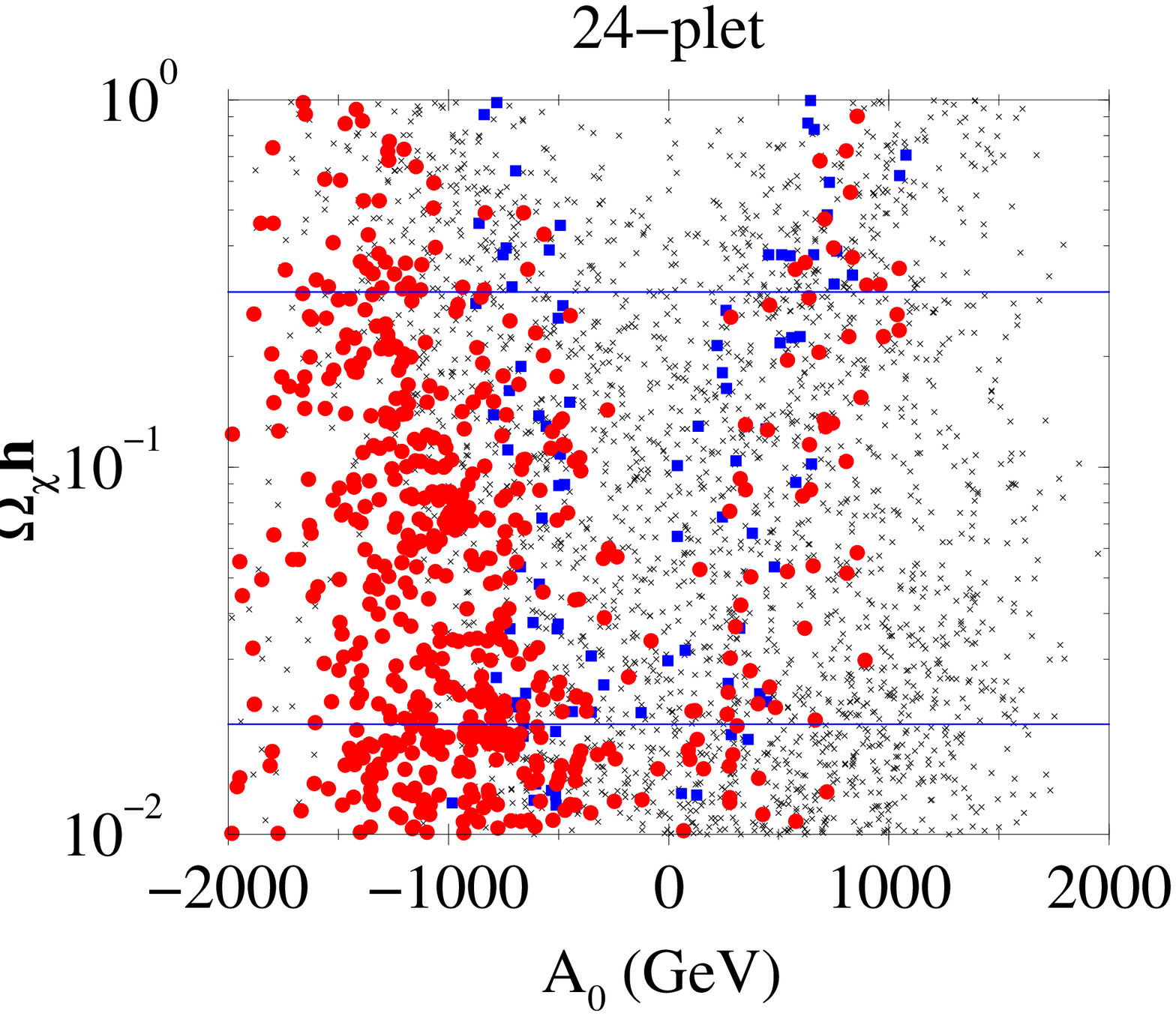}
\end{minipage}}                      
\hspace*{-0.6in}                      
\subfigure[]{    
\label{omega24e}                    
\begin{minipage}[b]{0.5\textwidth}                       
\centering                      
\includegraphics[width=\textwidth,height=0.65\textwidth]{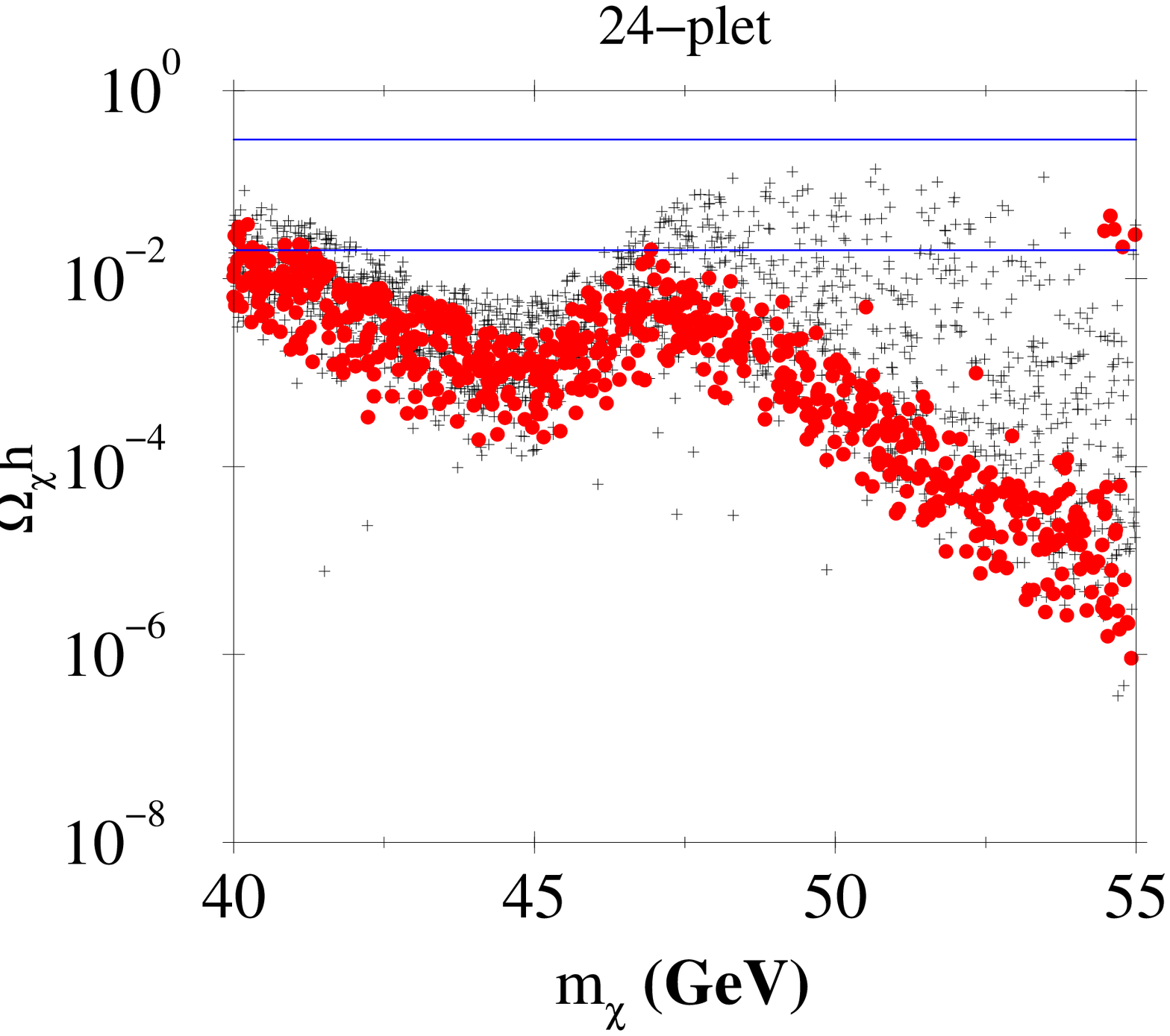}
\end{minipage}}                      
\hspace*{0.3in}
\subfigure[]{    
\label{omega24f}                    
\begin{minipage}[b]{0.5\textwidth}                       
\centering                      
\includegraphics[width=\textwidth,height=0.65\textwidth]{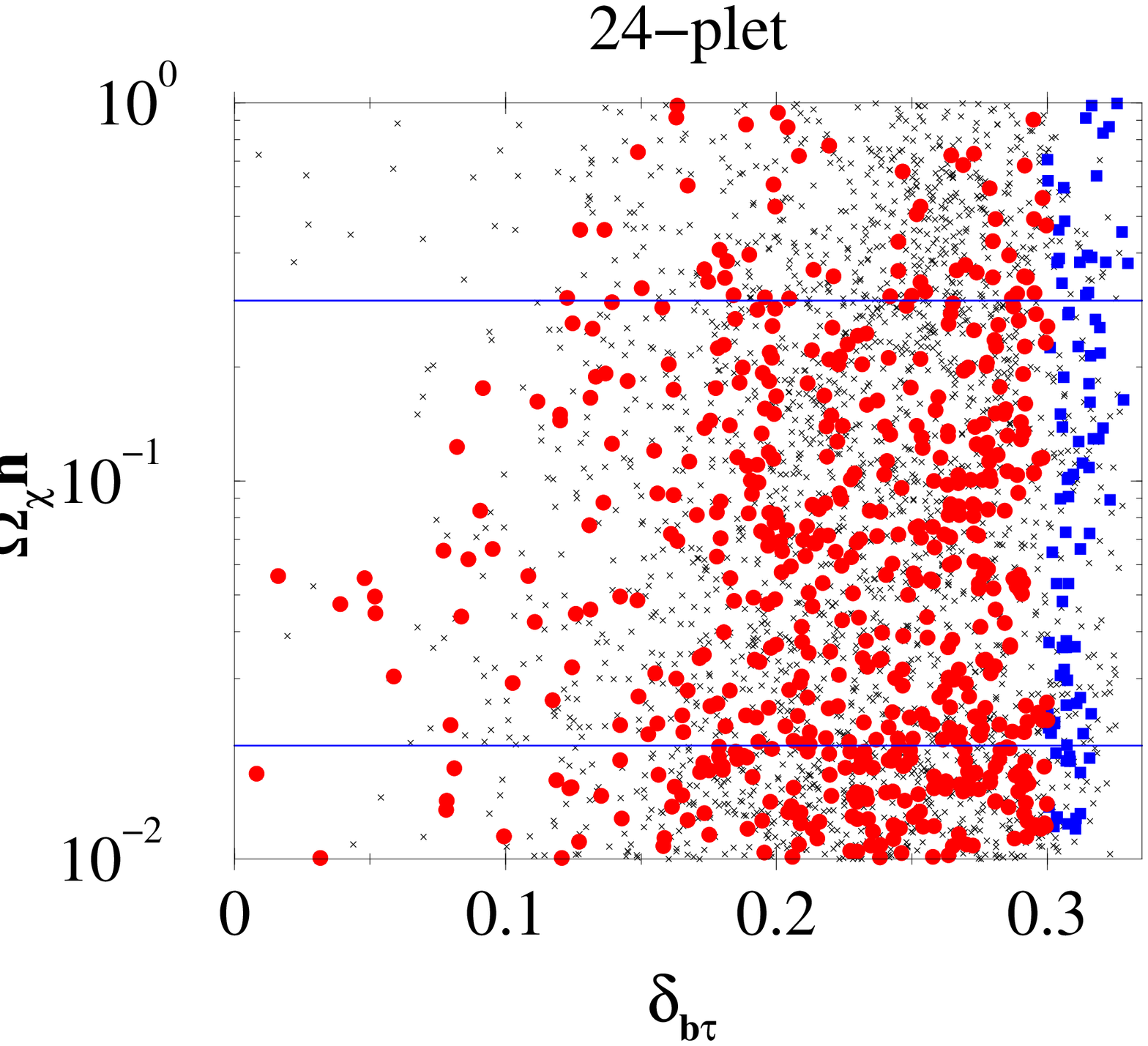}
\end{minipage}}                        
\caption{}                       
\label{omega24} 
\end{figure}

\noindent
Fig:~\ref{omega24}:\\
Fig.~\ref{omega24a}: Plot of $\Omegachi h^2$ vs $\tan\beta$ 
 for the SU(5) 24-plet case with the inputs  
$0<m_0<2$ TeV, $-1 ~{\rm TeV} <C_{24}m_{1/2}<1 ~{\rm TeV}$,
$-6 ~{\rm TeV}<A_0<6~{\rm TeV}$ and $\mu>0$. The small crosses 
satisfy the $g_{\mu}-2$ constraints, the (blue) filled squares 
additionally satisfy the $b \rightarrow s+ \gamma$ limits and the (red) filled 
ovals satisfy all the constraints, i.e.,  the $g_{\mu}-2$ constraint,
the $b \rightarrow s+ \gamma$ constraint, and 
$b-\tau$ unification at the level $\delta_{b\tau}\leq 0.3$.  
The two horizontal lines refer to the limits of Eq.(2)\\ 
Fig.~\ref{omega24b}: Plot of $\Omegachi h^2$  vs $m_0$ 
with all the same
parameters as in Fig.~\ref{omega24a} and
with $\tan\beta\leq 55$.
Symbols have the same meaning as in Fig.~\ref{omega24a}.\\
Fig.~\ref{omega24c}: Plot of $\Omegachi h^2$  vs 
$C_{24}*m_{\frac{1}{2}}$ with all other parameters the 
same as in Figs.~\ref{omega24a} and ~\ref{omega24b}.
Symbols have same meaning as in Fig.~\ref{omega24a}.\\
Fig.~\ref{omega24d}:  Plot of $\Omegachi h^2$  vs $A_0$ with all  other
parameters the same as in Figs.~\ref{omega24a}, ~\ref{omega24b} and 
~\ref{omega24c}.  Symbols have the same meaning as in Fig.~\ref{omega24a}.\\
Fig.~\ref{omega24e}:  Plot of $\Omegachi h^2$  vs the LSP 
mass $m_{\chi}$ with all other
parameters the same as in Figs.~\ref{omega24a} to ~\ref{omega24d}.
The small plus symbols refer to valid parameter points with no 
constraints, and 
(red) filled ovals refer to satisfying all the constraints 
i.e.,  the constraints from $g_{\mu}-2$, $b \rightarrow s+ \gamma$ and 
$\delta_{b\tau}\leq 0.3$.\\
Fig.~\ref{omega24f}:  Plot of $\Omegachi h^2$  vs 
$\delta_{b\tau}$ with all other
parameters the same as in Figs.~\ref{omega24a} to ~\ref{omega24d}.
Symbols have the same meaning as in Fig.~\ref{omega24a}.\\

\begin{figure}           
\vspace*{-2.0in}                                 
\subfigure[]{                       
\label{omega54a} 
\hspace*{-0.6in}                     
\begin{minipage}[b]{0.5\textwidth}                       
\centering
\includegraphics[width=\textwidth,height=0.65\textwidth]{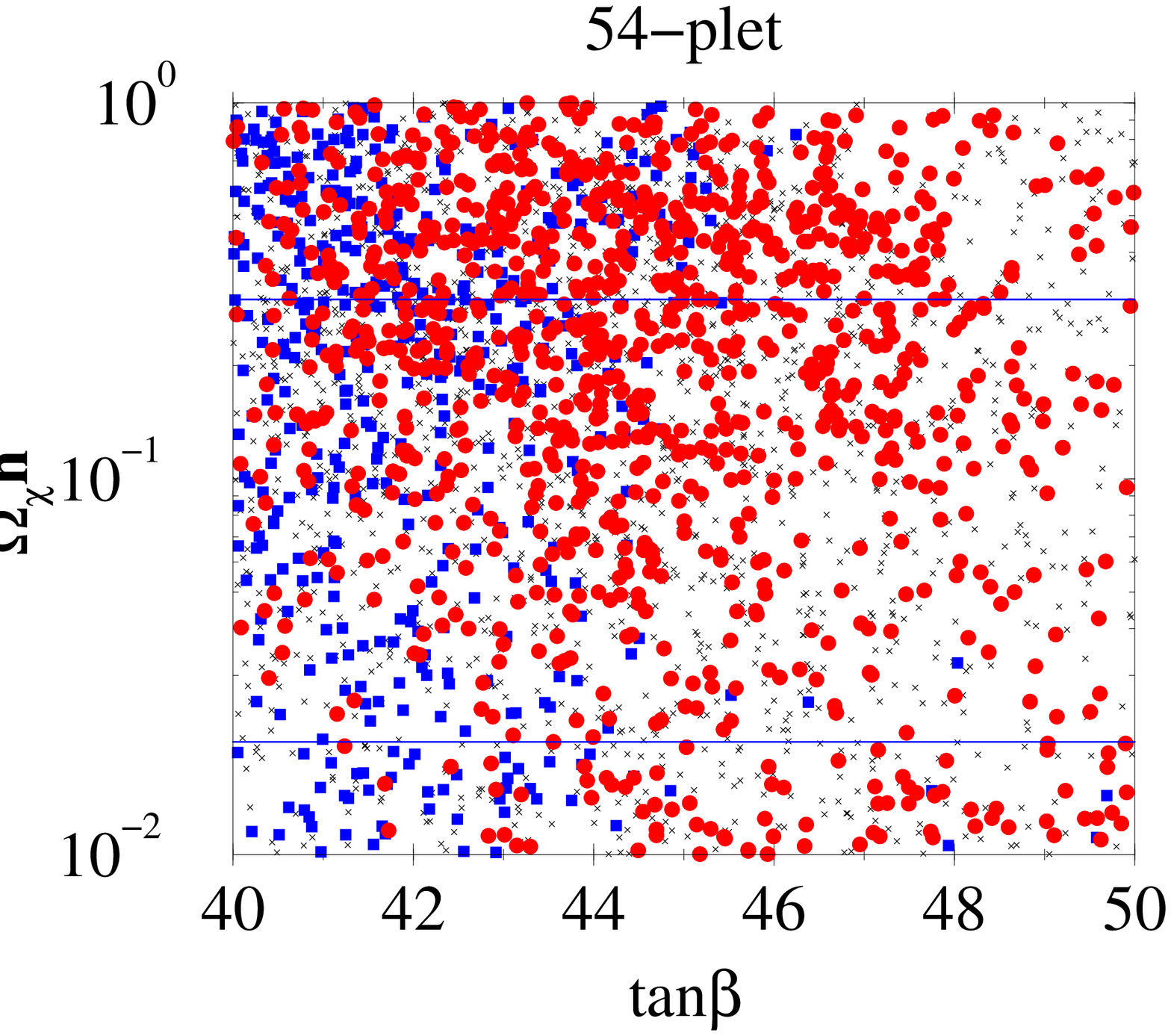}
\end{minipage}}                       
\hspace*{0.3in}
\subfigure[]{    
\label{omega54b}   
\begin{minipage}[b]{0.5\textwidth}                       
\centering                      
\includegraphics[width=\textwidth,height=0.65\textwidth]{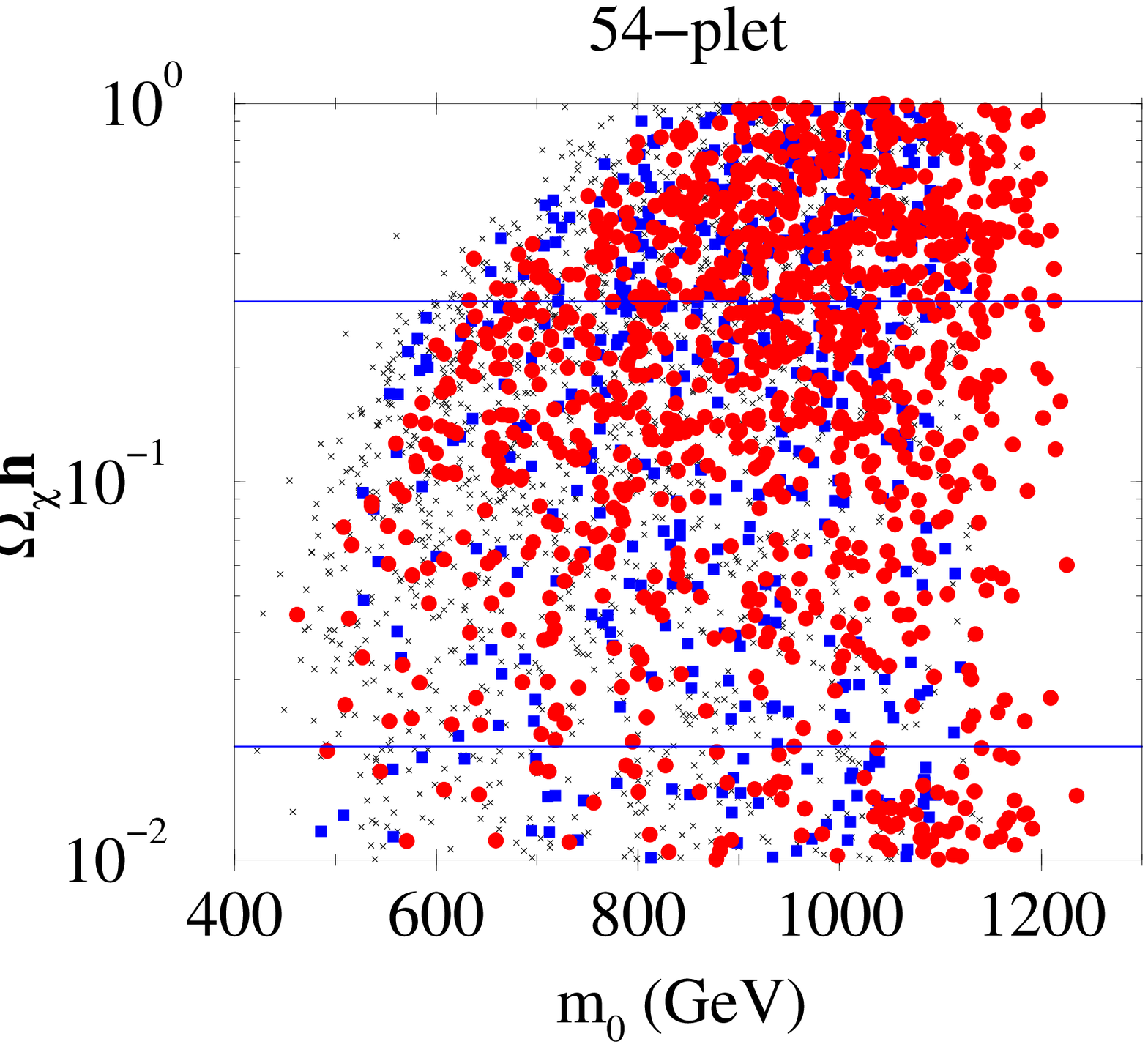} 
\end{minipage}}                       
\hspace*{-0.6in}                     
\subfigure[]{    
\label{omega54c}                    
\begin{minipage}[b]{0.5\textwidth}                       
\centering
\includegraphics[width=\textwidth,height=0.65\textwidth]{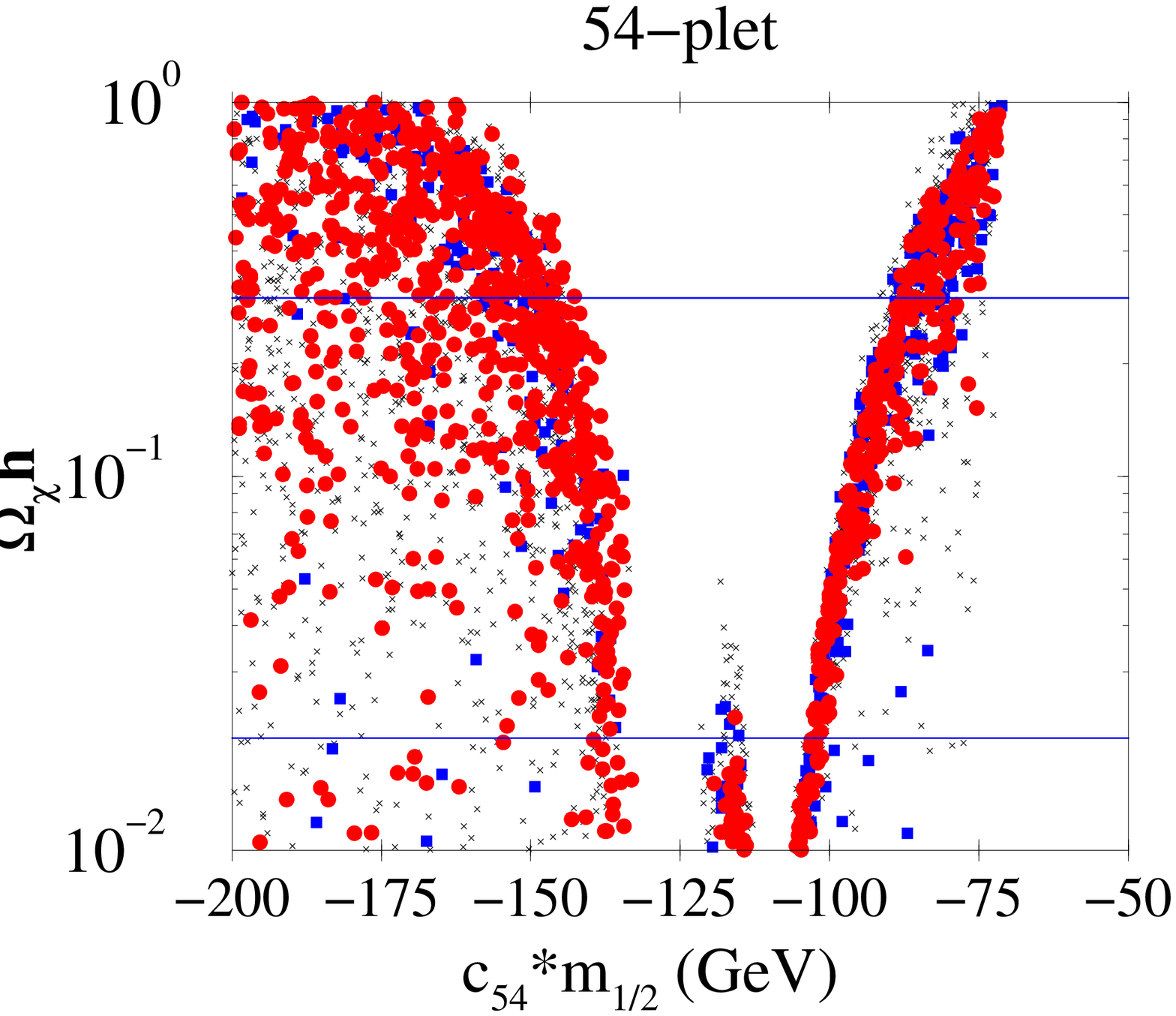}
\end{minipage}}
\hspace*{0.3in}                       
\subfigure[]{    
\label{omega54d}                    
\begin{minipage}[b]{0.5\textwidth}                       
\centering                      
\includegraphics[width=\textwidth,height=0.65\textwidth]{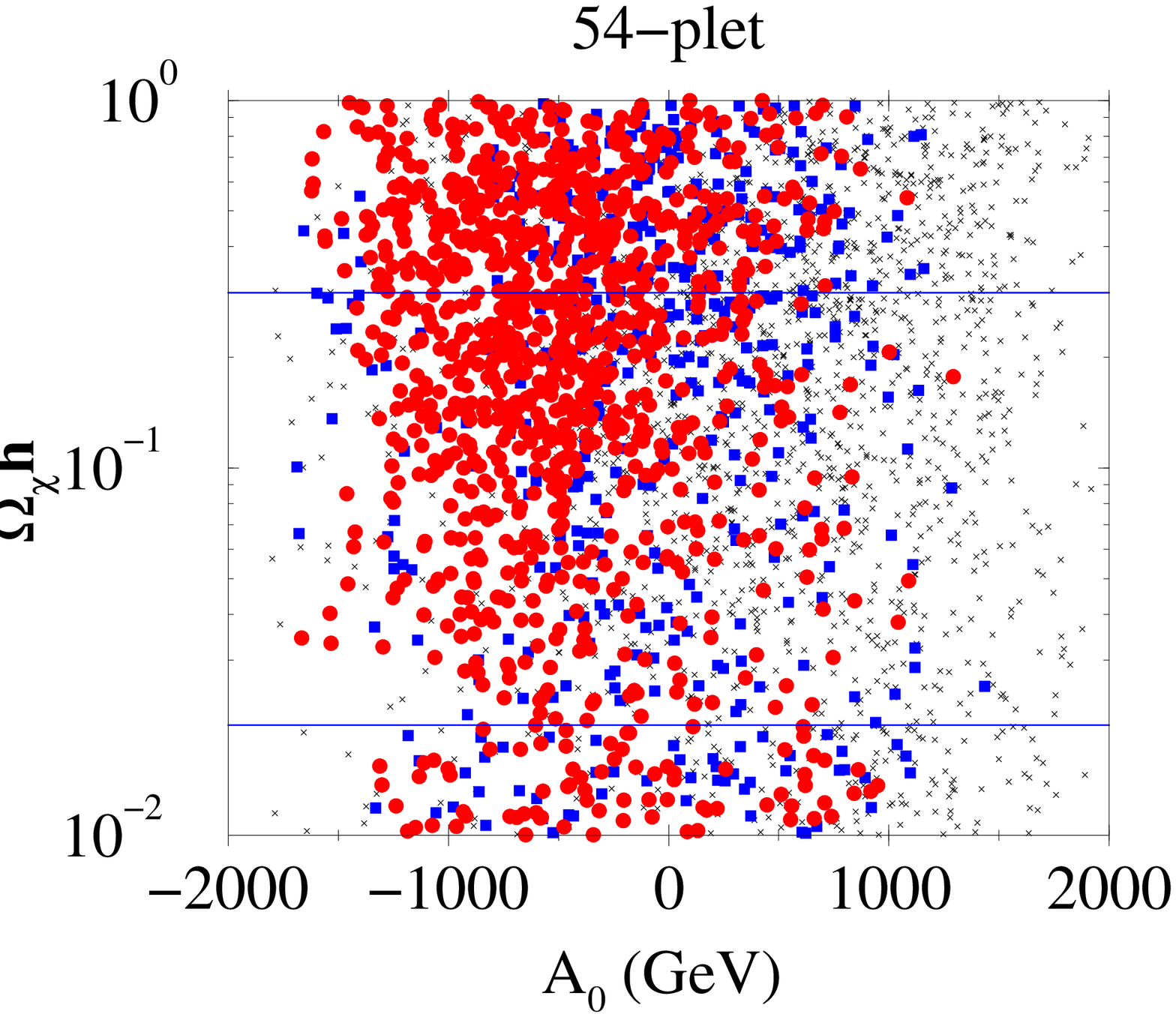}
\end{minipage}}
\hspace*{-0.6in}
\subfigure[]{    
\label{omega54e}                    
\begin{minipage}[b]{0.5\textwidth}                       
\centering                      
\includegraphics[width=\textwidth,height=0.65\textwidth]{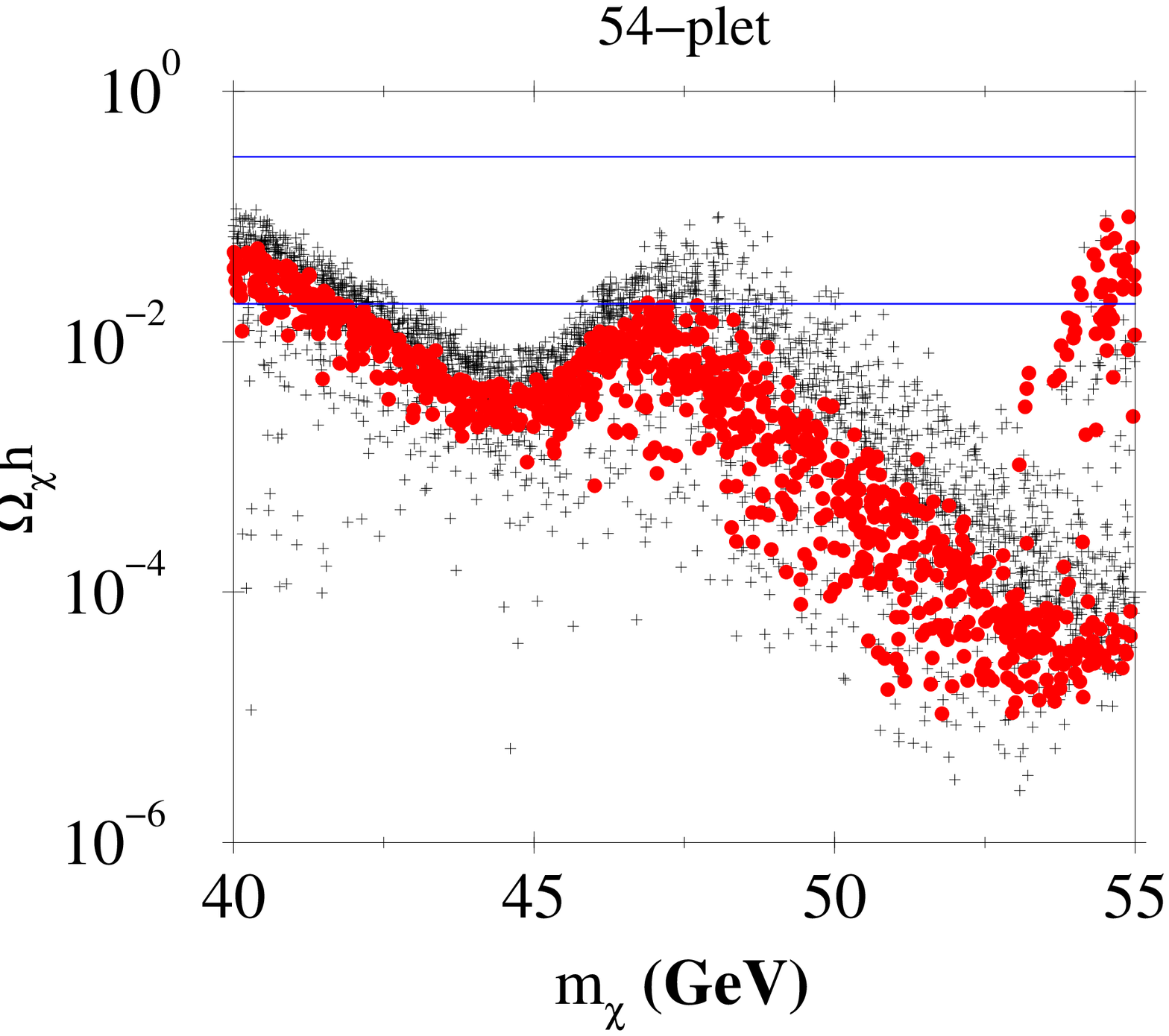}
\end{minipage}}
\hspace*{0.3in}                       
\subfigure[]{    
\label{omega54f}                    
\begin{minipage}[b]{0.5\textwidth}                       
\centering                      
\includegraphics[width=\textwidth,height=0.65\textwidth]{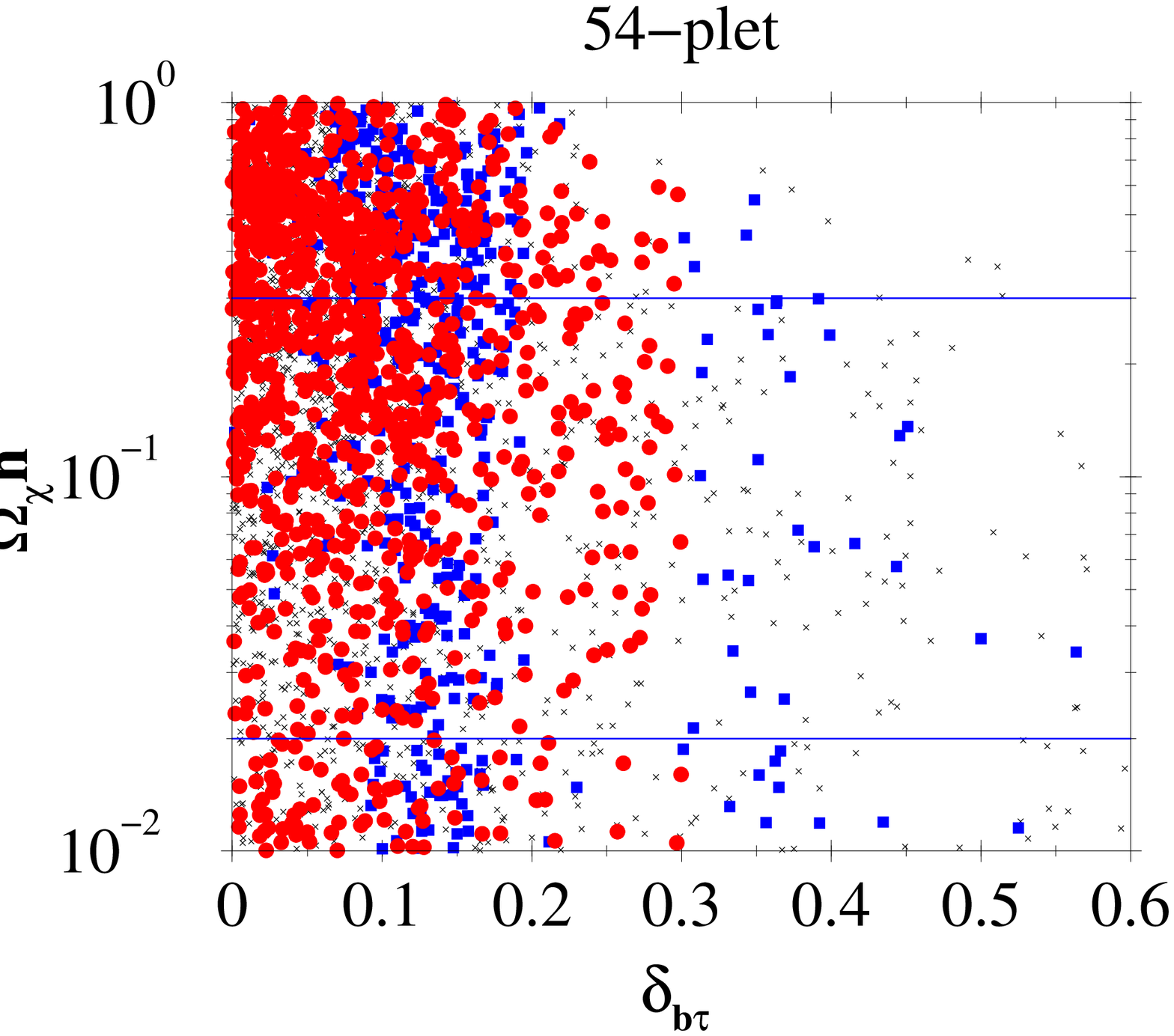}
\end{minipage}}
\caption{}                       
\label{omega54} 
\end{figure}

\noindent
Fig:~\ref{omega54}:\\
Fig.~\ref{omega54a}: Plot of $\Omegachi h^2$ vs $\tan\beta$ 
for the SO(10) 54-plet case with inputs
$0<m_0<2$ TeV, $-1 ~{\rm TeV} <C_{54}m_{1/2}<1 ~{\rm TeV}$,
$-6 ~{\rm TeV}<A_0<6~{\rm TeV}$ and $\mu>0$. The nonuniversal Higgs scalar 
parameters are given by $m_{H_1}^2=1.5m_0^2$ and $m_{H_2}^2=0.5m_0^2$.  
The small crosses 
satisfy the $g_{\mu}-2$ constraints, the (blue) filled squares 
additionally satisfy the $b \rightarrow s+ \gamma$ limits and the (red) filled 
ovals satisfy all the constraints, i.e.,  the $g_{\mu}-2$ constraint,
the $b \rightarrow s+ \gamma$ constraint, and 
unification of Yukawa couplings so that 
$\delta_{b\tau},\delta_{bt}, \delta_{t\tau} \leq 0.3$.  
The two horizontal lines refer to the limits of Eq.(2)\\
Fig.~\ref{omega54b}: Plot of $\Omegachi h^2$  vs $m_0$ 
with all the same
parameters as in Fig.~\ref{omega54a} and with $\tan\beta\leq 55$.
Symbols have the same meaning as in Fig.~\ref{omega54a}.\\
Fig.~\ref{omega54c}: Plot of $\Omegachi h^2$  vs 
$C_{54}*m_{\frac{1}{2}}$ with all other parameters the 
same as in Figs.~\ref{omega54a} and ~\ref{omega54b}.
Symbols have the same meaning as in Fig.~\ref{omega54a}.\\
Fig.~\ref{omega54d}:  Plot of $\Omegachi h^2$  vs $A_0$ with all  other
parameters the same as in Figs.~\ref{omega54a}, ~\ref{omega54b} and 
~\ref{omega54c}.  Symbols have the same meaning as in Fig.~\ref{omega54a}.\\
Fig.~\ref{omega54e}:  Plot of $\Omegachi h^2$  vs the LSP 
mass $m_{\chi}$ with all other
parameters the same as in Figs.~\ref{omega54a} to ~\ref{omega54d}.
The small plus symbols refer valid parameter points with 
no constraints, and (red) filled ovals refer to satisfying all the constraints 
i.e.,  the constraints from $g_{\mu}-2$, $b \rightarrow s+ \gamma$ and 
$\delta_{b\tau},\delta_{bt} \delta_{t\tau} \leq 0.3$.\\
Fig.~\ref{omega54f}:  Plot of $\Omegachi h^2$  vs 
$\delta_{b\tau}$ with all other
parameters the same as in Figs.~\ref{omega54a} to ~\ref{omega54d}.
Symbols have the same meaning as in Fig.~\ref{omega54a}.\\

\begin{figure}           
\vspace*{-2.0in}                                 
\subfigure[]{                       
\label{sigma24mchi} 
\hspace*{-0.6in}                     
\begin{minipage}[b]{0.5\textwidth}                       
\centering
\includegraphics[width=\textwidth,height=0.65\textwidth]{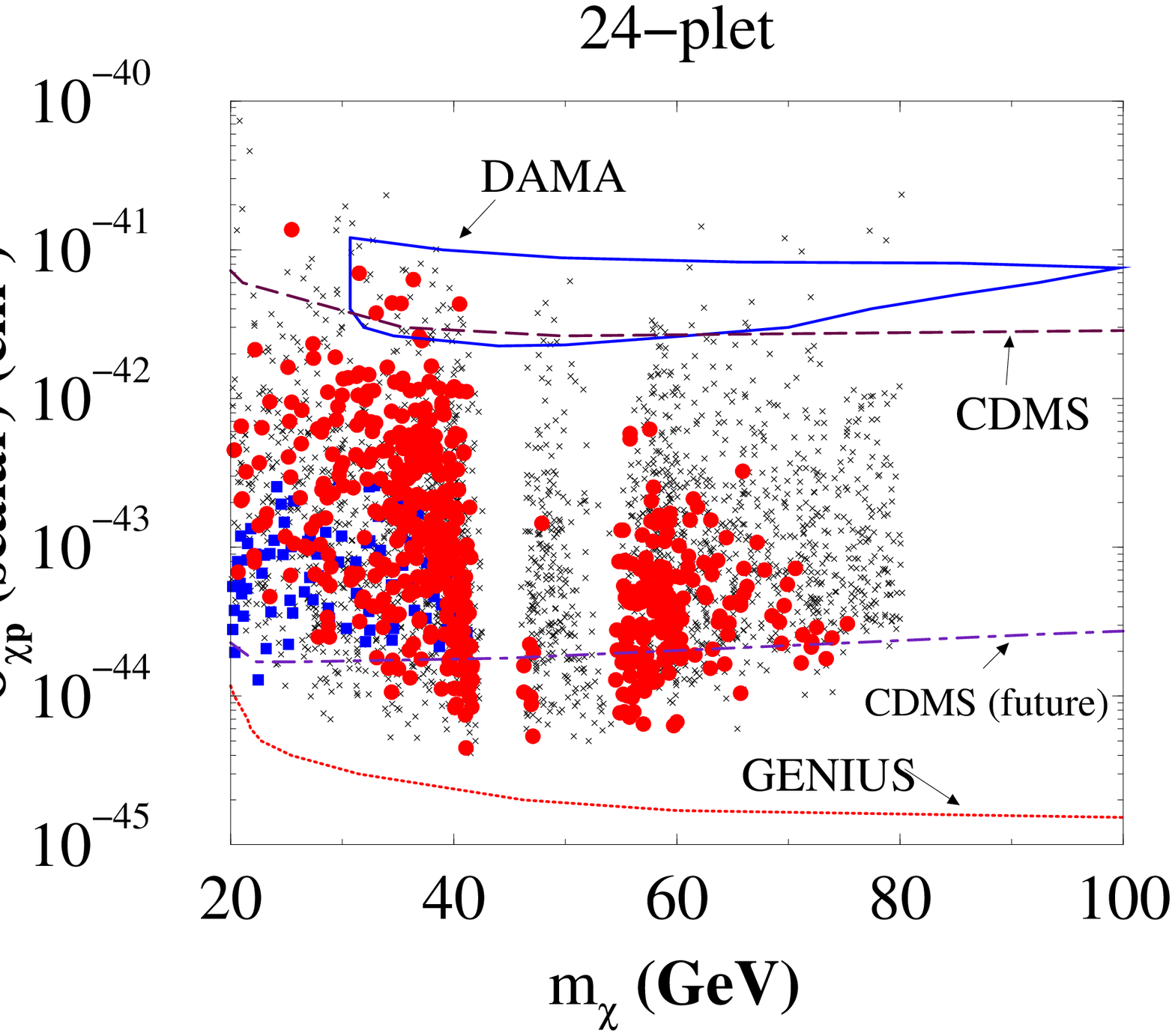}
\end{minipage}}                       
\hspace*{0.3in}
\subfigure[]{    
\label{sigma24tan}
\begin{minipage}[b]{0.5\textwidth}                       
\centering                      
\includegraphics[width=\textwidth,height=0.65\textwidth]{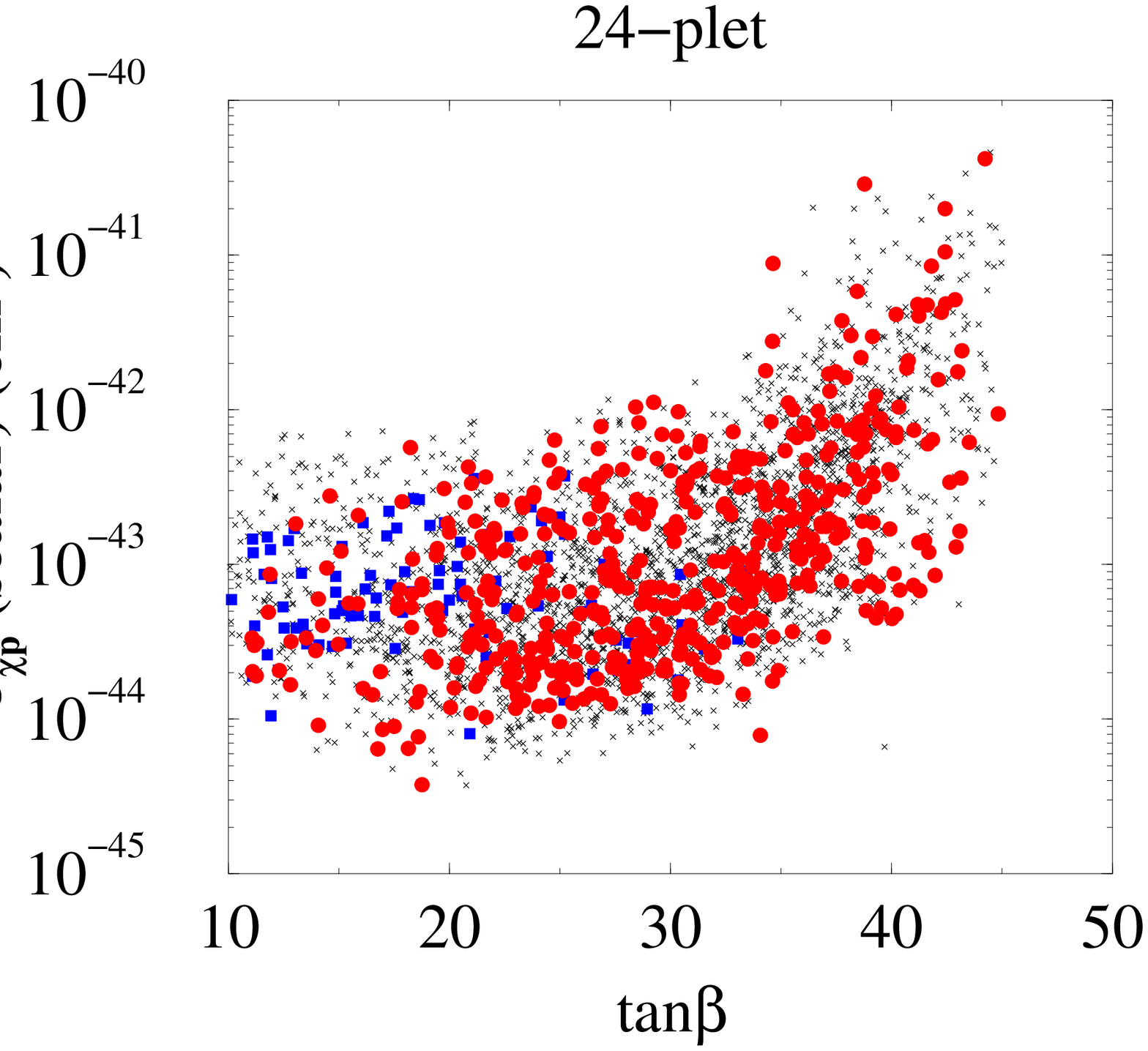}
\end{minipage}}                       
\hspace*{-0.6in}                     
\subfigure[]{    
\label{sigma54mchi}                     
\begin{minipage}[b]{0.5\textwidth}                       
\centering
\includegraphics[width=\textwidth,height=0.65\textwidth]{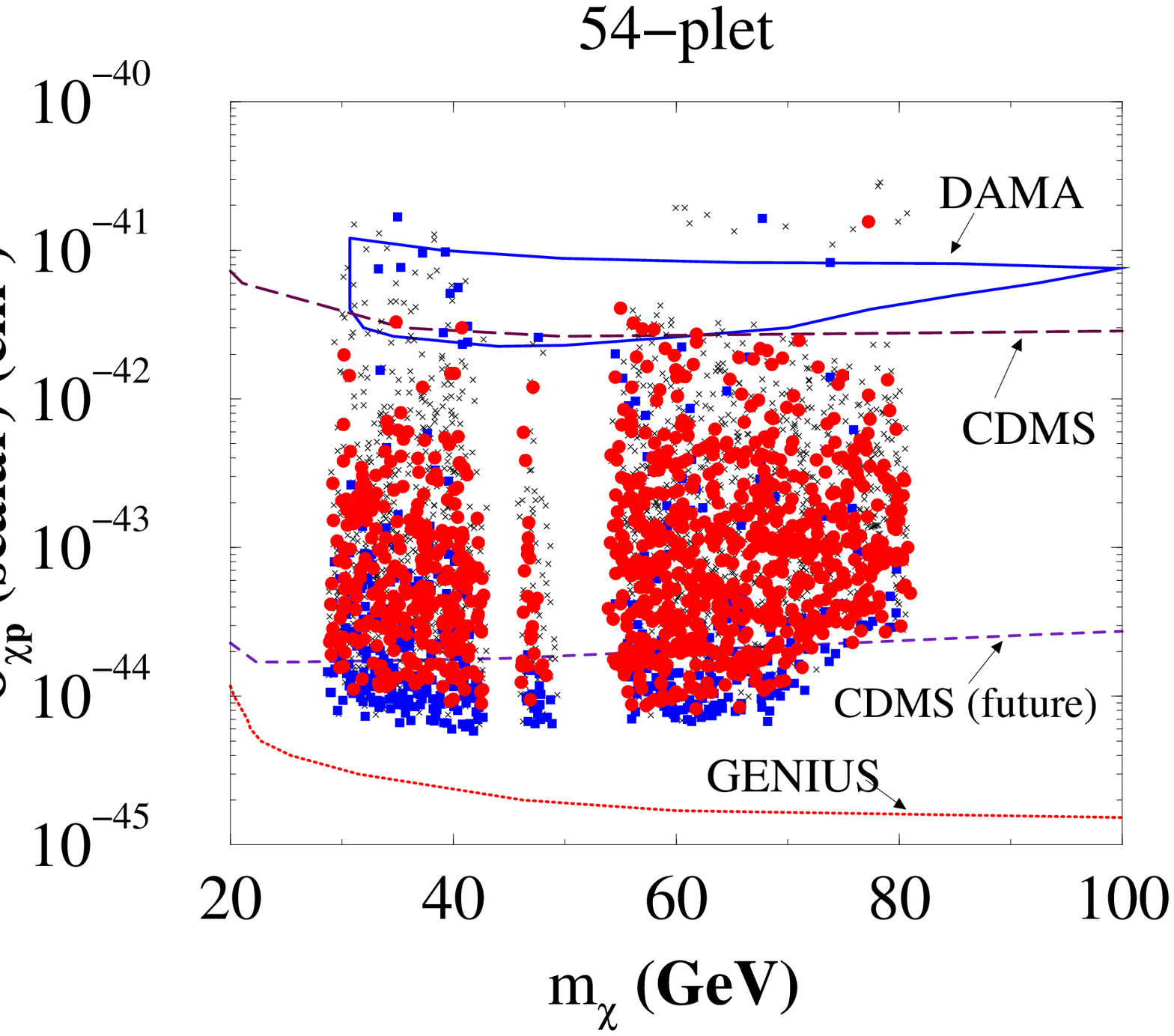}
\end{minipage}}
\hspace*{0.3in}                       
\subfigure[]{    
\label{sigma54tan}                   
\begin{minipage}[b]{0.5\textwidth}                       
\centering                      
\includegraphics[width=\textwidth,height=0.65\textwidth]{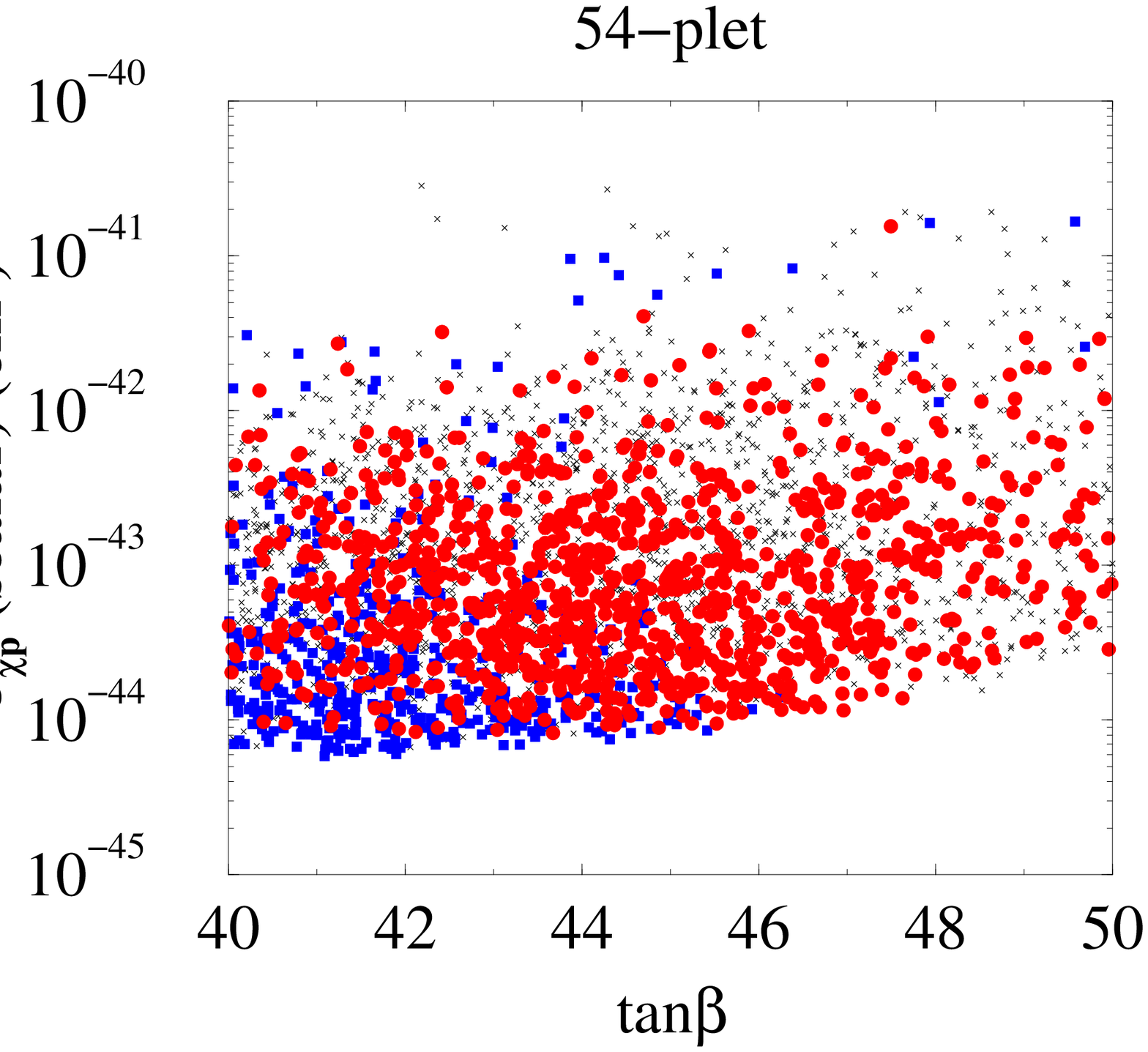}
\end{minipage}}                       
\hspace*{-0.6in}                       
\subfigure[]{    
\label{sigma54primemchi}                   
\begin{minipage}[b]{0.5\textwidth}                       
\centering                      
\includegraphics[width=\textwidth,height=0.65\textwidth]{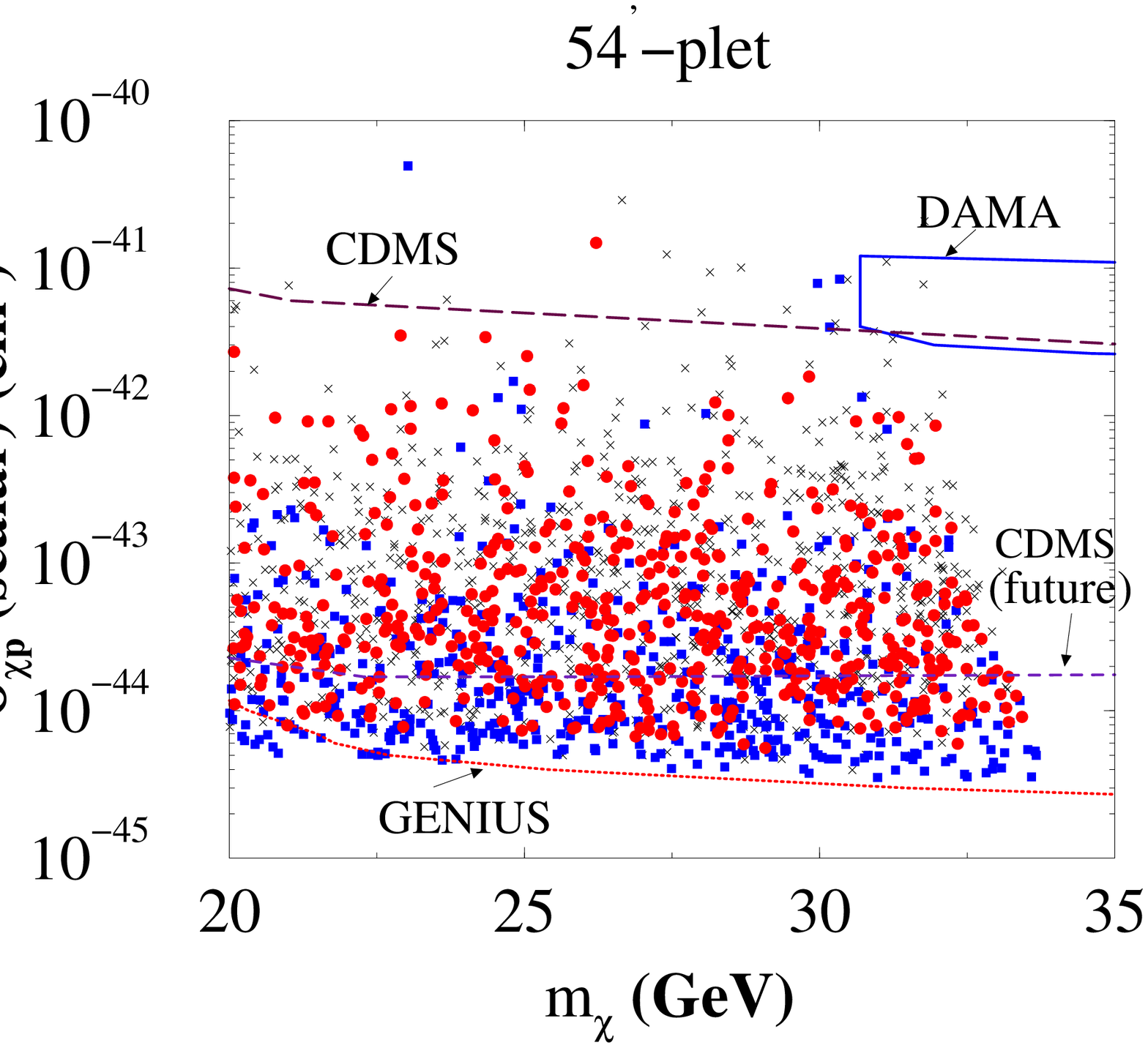}
\end{minipage}}                       
\hspace*{0.3in}
\subfigure[]{
\label{sigma54primetan}
\begin{minipage}[b]{0.5\textwidth}
\centering
\includegraphics[width=\textwidth,height=0.65\textwidth]{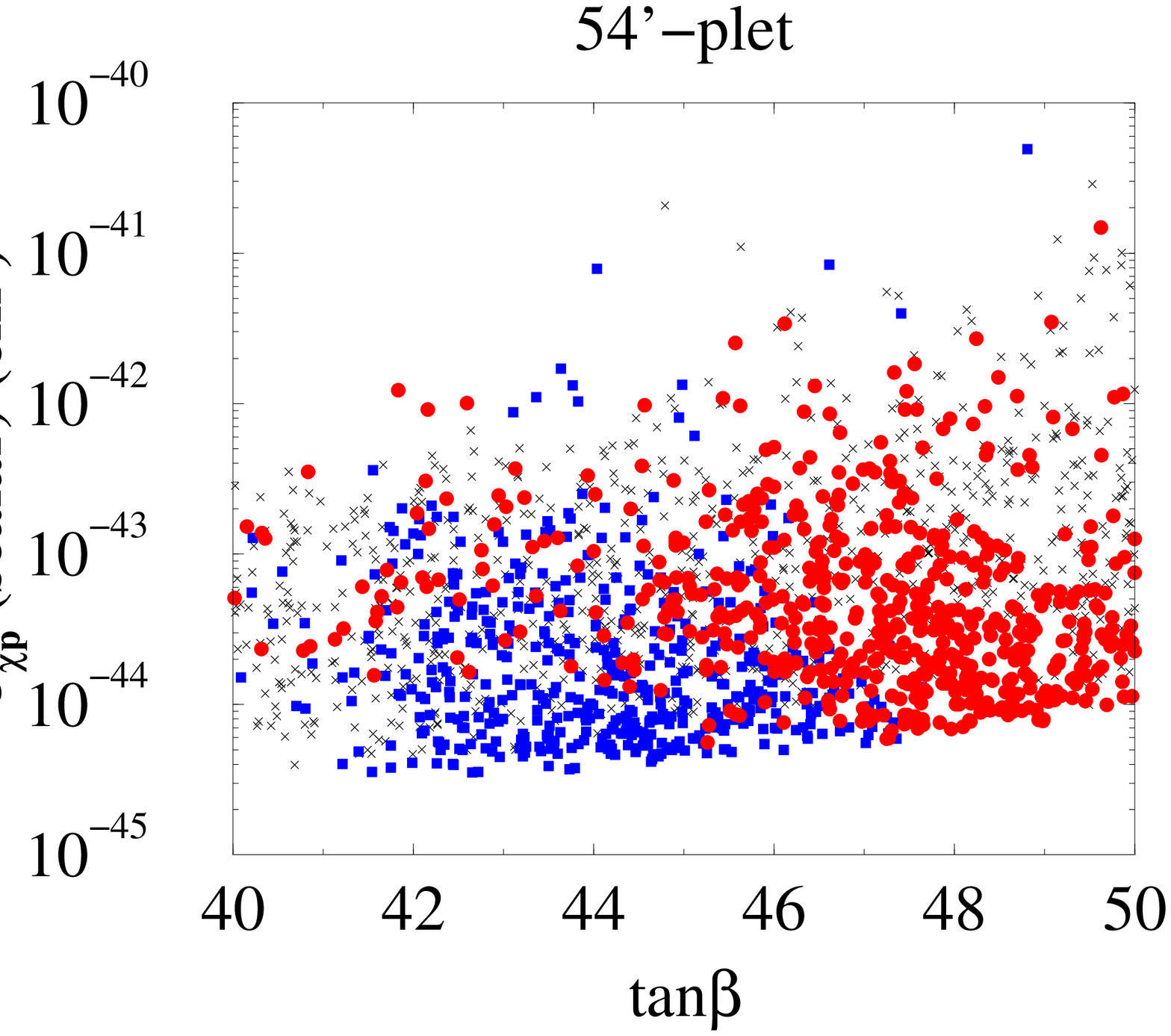}
\end{minipage}}
\caption{}                       
\label{omegasigma}
\end{figure}

\noindent
Fig.~\ref{omegasigma}:\\
Fig.~\ref{sigma24mchi}: Plot of the neutralino-proton scalar 
cross section $\sigma_{\chi p}$ vs the lightest neutralino 
mass $m_{\chi}$ for the SU(5) 24 plet case with the range of the 
parameters given in Figs.~\ref{omega24a} to ~\ref{omega24d} 
satisfying all the desired constraints including
the $b-\tau$ unification constraint so that $\delta_{b\tau}\leq 0.3$.
The small crosses 
satisfy the $g_{\mu}-2$ constraints, the (blue) filled squares 
additionally satisfy the $b \rightarrow s+ \gamma$ limits and the (red) filled 
ovals satisfy all the constraints, i.e.,  the $g_{\mu}-2$ constraint,
the $b \rightarrow s+ \gamma$ constraint, and $\delta_{b\tau}\leq 0.3$.
The area enclosed by solid lines is excluded by the DAMA 
experiment\cite{dama}, the dashed line is the lower limit from the
CDMS experiment\cite{cdms}, the dot-dashed line is the lower limit 
achievable by CDMS in the future\cite{cdms} and the dotted line is the 
lower limit expected from the proposed GENIUS experiment\cite{genius}.\\
Fig.~\ref{sigma24tan}: Plot of the neutralino-proton scalar 
cross section $\sigma_{\chi p}$ vs $\tanbeta$ for the 
SU(5) 24 plet case with the same range of parameters as given in
Figs.~\ref{omega24a} - ~\ref{omega24d} satisfying all 
the desired constraints including
the $b-\tau$ unification constraint so that $\delta_{b\tau}\leq 0.3$.
Symbols have the same meaning as in Fig.~\ref{sigma24mchi}. \\
Fig.~\ref{sigma54mchi}: Plot of the neutralino-proton scalar 
cross section $\sigma_{\chi p}$ vs the neutralino mass $m_{\chi}$ for the 
SO(10) 54 plet case with the same range of parameters as given in
Figs.~\ref{omega54a} - ~\ref{omega54d} 
satisfying all the desired constraints including
the $b-\tau$, $b-t$ and  $t-\tau$ unification constraint so that 
$\delta_{b\tau}, \delta_{bt}, \delta_{t\tau}\leq 0.3$. The small crosses 
satisfy the $g_{\mu}-2$ constraints, the (blue) filled squares 
additionally satisfy the $b \rightarrow s+ \gamma$ limits and the (red) filled 
ovals satisfy all the constraints, i.e.,  the $g_{\mu}-2$ constraint,
the $b \rightarrow s+ \gamma$ constraint, and 
Yukawa unifications with $\delta_{b\tau},\delta_{bt} \delta_{t\tau} 
\leq 0.3$.\\
Fig.~\ref{sigma54tan}: Plot of the neutralino-proton scalar 
cross section $\sigma_{\chi p}$ vs $\tanbeta$ for the 
SO(10) 54 plet case with the same range of parameters as given in
Figs.~\ref{omega54a} - ~\ref{omega54d} 
satisfying all the desired constraints including
the $b-\tau$, $b-t$ and  $t-\tau$ unification constraint so that 
$\delta_{b\tau}, \delta_{bt}, \delta_{t\tau}\leq 0.3$.
Symbols have the same meaning as in Fig.~\ref{sigma54mchi}.\\ 
Fig.~\ref{sigma54primemchi}: Plot of the neutralino-proton scalar 
cross section $\sigma_{\chi p}$ vs the neutralino mass $m_{\chi}$ for the 
SO(10) $54'$ plet case with the same range of parameters as given in
Figs.~\ref{omega54a} - ~\ref{omega54d} 
satisfying all the desired constraints including
the $b-\tau$, $b-t$ and  $t-\tau$ unification constraint so that 
$\delta_{b\tau}, \delta_{bt}, \delta_{t\tau}\leq 0.3$.
Symbols have the same meaning as in Fig.~\ref{sigma54mchi}. \\
Fig.~\ref{sigma54primetan}: Plot of the neutralino-proton scalar 
cross section $\sigma_{\chi p}$ vs $\tanbeta$ for the 
SO(10) $54'$ plet case with the same range of parameters as given in
Figs.~\ref{omega54a} - ~\ref{omega54d} 
satisfying all the desired constraints including
the $b-\tau$, $b-t$ and  $t-\tau$ unification constraint so that 
$\delta_{b\tau}, \delta_{bt}, \delta_{t\tau}\leq 0.3$.
Symbols have the same  meaning as in Fig.~\ref{sigma54mchi}. \\

\newpage

\newpage


\newpage

\newpage

\newpage

\end{document}